\documentclass[11pt]{article}
\usepackage{amsmath, amsfonts, bm, amssymb}
\usepackage{amsfonts, graphicx, grffile, float, latexsym, mathtools}
\usepackage{dsfont}
\usepackage{multicol}
\usepackage{comment}
\usepackage{cite}
\usepackage{tikz}
\usetikzlibrary{positioning}
\usetikzlibrary{calc}
\usepackage{wrapfig}
\usepackage{mathrsfs}
\makeatletter
\let\NAT@parse\undefined
\makeatother
\usepackage{amsthm}
\usepackage{bbm}
\usepackage{bm}

\makeatletter
\renewcommand*\env@matrix[1][c]{\hskip -\arraycolsep
  \let\@ifnextchar\new@ifnextchar
  \array{*\c@MaxMatrixCols #1}}

\usepackage{Shorthands}
\usepackage{enumitem}

\DeclareMathOperator{\diag2}{diag}

\makeatletter
\let\NAT@parse\undefined
\makeatother
\usepackage{hyperref}
\hypersetup{%
    pdfborder = {0 0 0}
}

\usepackage{cleveref}
\crefname{assumption}{}{}
\newtheoremstyle{named}{}{}{\itshape}{}{\bfseries}{.}{.5em}{\thmnote{#3}}
\theoremstyle{named}

\newcommand{\Acl}{A_K^{(2)}}

\newcommand{\E}{\ensuremath{\mathbf{E}}}
\newcommand{\hatx}{\hat{x}}
\usepackage{stfloats}
\usepackage{subcaption}

\usepackage{algorithm} 
\usepackage{algpseudocode} 
\usepackage{authblk}

\usepackage{caption}

\usepackage{xcolor}
\definecolor{teal}{HTML}{57C5D4}
\definecolor{mustard}{HTML}{FAC17F}
\usepackage{fullpage}

\title{Layered Control of Partially Observed Stochastic Systems}

\author[1]{Charis Stamouli}
\author[2]{Anastasios Tsiamis}
\author[1]{George J. Pappas}

\affil[1]{GRASP Lab, University of Pennsylvania}
\affil[2]{Automatic Control Laboratory, ETH Zürich}

\date{}

\begin{document}

\maketitle
\thispagestyle{empty}
\pagestyle{empty}
\captionsetup[figure]{labelfont={bf},labelformat={default},labelsep=period,name={Fig.}}

\begin{abstract}
Layered control is essential for managing complexity in large-scale systems, employing progressively coarser models at higher layers. While significant advances have been made for fully observable systems, the  theoretical foundations of layered control  under partial observations and stochastic noise remain underexplored. To address this gap, we propose a principled layered control framework for such settings. Given a state estimator at each layer, our approach ensures that the expected output distance between systems at successive layers remains within a priori computable bounds. This is achieved by introducing a novel notion of stochastic simulation functions for partially observed systems. For the class of linear systems with Kalman estimators, we provide a systematic construction of these functions along with the corresponding control design. We demonstrate our framework on two aerial robotic scenarios: an unmanned aerial vehicle and a hexacopter with a camera payload.
\end{abstract}

\section{Introduction}
Layered control is a powerful framework for managing complexity in large-scale systems~\cite{matni2024towards}. The key idea is to employ progressively coarser models of the system at higher layers. While significant efforts have been devoted to advancing the design of individual layers, their interplay is typically handled through ad-hoc engineering. A fundamental challenge that arises in this context is ensuring that lower layers can implement the commands of higher layers. This is essential when layered architectures are deployed in safety-critical domains, including flight control \cite{tischler2005modernized}, chemical process control \cite{seborg2016process}, and unmanned aerial vehicle (UAV) control \cite{beard2012small}. 

Principled layered control approaches have been developed  to ensure that lower layers can approximate the behavior of higher layers  within  computable bounds \cite{Girard2009, Fainekos2009}. Smaller  bounds may suggest a good architectural design, whereas larger ones may imply the need for redesigning the architecture. In \cite{Girard2009, Fainekos2009}, the control design accounts for fully observable systems without stochastic noise. However, partial observations and measurement noise often arise in practice, introducing further challenges for layered control. %Examples of relevant applications include ..., ...., and ... . The fundamental problem 

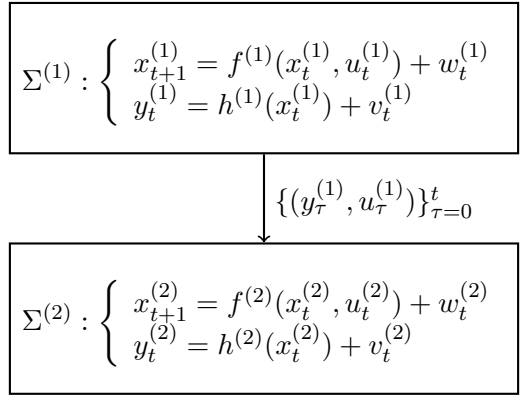
\begin{figure}[tbh]
\begin{center}
    \begin{tikzpicture}
        % Top rectangle
        \node[rounded corners=8pt,fill=teal!20, thick, minimum width=4cm, minimum height=2cm] (top) at (0,2) {
            \begin{minipage}{6.5cm}
                \centering
                \vspace{-0.3cm}
                \begin{align*}
                    \hspace{.0cm}\Sigma^{(1)}:\left\{
                        \begin{array}{lcr}
                            x_{t+1}^{(1)}=f^{(1)}(x_t^{(1)},u_t^{(1)})+w_t^{(1)} \\
                            y_t^{(1)} = h^{(1)}(x_t^{(1)})+v_t^{(1)}
                        \end{array}
                        \right. 
                \end{align*}
            \end{minipage}
        };
        
        % Bottom rectangle
        \node[rounded corners=8pt,fill=mustard!30, thick, minimum width=4cm, minimum height=2cm] (bottom) at (0, -1.2) {
            \begin{minipage}{6.5cm}
                \centering
                \vspace{-0.3cm}
                \begin{align*}
                        \hspace{.0cm}\Sigma^{(2)}:\left\{
                        \begin{array}{lcr}
                            x_{t+1}^{(2)}=f^{(2)}(x_t^{(2)},u_t^{(2)})+w_t^{(2)} \\
                            y_t^{(2)} = h^{(2)}(x_t^{(2)})+v_t^{(2)}
                        \end{array}
                        \right.
                \end{align*}
            \end{minipage}
        };
        
        % Shorter arrow with text
        \draw[->, thick] (top.south) -- (bottom.north) node[midway, right] {$\{(y_\tau^{(1)},u_{\tau}^{(1)})\}_{\tau=0}^t$};
        % Dashed arrow looping back
        %\draw[dashed, ->] (bottom.south) -- ++(0,-0.5) -- ++(4,0) -- ++(0,7.1) node[midway, left] {$\barX_+,\;\barU_+$} -- ++(-4,0) -- (top.north);
    \end{tikzpicture}
\end{center}
 \caption{Two-layer control architecture for partially observed stochastic systems. Our goal is to control system $\Sigma^{(2)}$ so that it approximates the output behavior of $\Sigma^{(1)}$ under any controller within guaranteed computable  bounds.
 }
\label{fig:problem_architecture}
\end{figure}

In this paper, we present a principled framework for layered control of partially observed stochastic systems (see Fig.~\ref{fig:problem_architecture}). Given access to a state estimator for each system, we  define a lower-layer controller inspired by stochastic simulation functions \cite{julius2009approximations}. Simulation functions are control Lyapunov-like functions that describe how the lower-layer system can approximate any closed-loop behavior of the higher-layer system within guaranteed bounds. Our main contributions are as follows:
\begin{itemize}
\item We propose a principled framework for layered control of partially observed stochastic systems.
\item Our framework relies on novel stochastic simulation functions tailored to partially observed stochastic systems equipped with state estimators.
\item We prove that our method guarantees bounded expected distance between the systems over time. The distance bound can be computed a priori to assess whether the layered control architecture is well designed.
\item For the class of linear systems with Kalman estimators, we provide a systematic construction of the stochastic simulation functions and the corresponding control design.
\end{itemize}

Our results focus on two-layer architectures, but our approach readily applies to additional layers.
We demonstrate our framework on two aerial robotic scenarios. In the first scenario, a UAV prototype serves as the higher-layer system, while the same UAV with added control surfaces is the lower-layer system. In the second scenario, a quadcopter is the higher-layer system and a hexacopter with a camera is the lower-layer system. In both scenarios, the lower-layer system closely matches the behavior of the higher-layer system. We also observe that our output distance bound is quite tight relative to the empirical value across Monte Carlo trials.

All proofs are provided in Appendix~\ref{app:Proofs}.
\medskip

\noindent{\bf Related work.} The problem of layered control has been extensively studied in the literature. Previous works focus primarily on systems with full-state observations, either in noiseless \cite{matni2016theory,rosolia2022unified,Girard2009} or in stochastic settings \cite{cominesi2017two,van2024specification,lavaei2017compositional}. Among these works, the approaches in \cite{Girard2009,lavaei2017compositional} provide formal guarantees of bounded output distance between the systems composing the architecture. In the former, this is achieved by introducing simulation functions for non-stochastic systems, while the latter relies on a notion of stochastic pseudo-simulation functions.   

Partial observability has been considered in related but distinct settings~\cite{lesser2015controller,haesaert2015observer}. In particular, the work in~\cite{lesser2015controller} considers observer-based control of a \emph{single} partially observed stochastic system, bounding the distance between true states and their estimates. In parallel, the work in~\cite{haesaert2015observer} designs an interface controller between a partially observed stochastic linear system and a noiseless counterpart based on simulation relations. In contrast, we consider \emph{two} partially observed stochastic systems across layers, allowing for \emph{states and inputs with distinct dimensions and physical interpretations}. This setting arises naturally when a controller has been designed and validated on one plant---a scaled prototype, a laboratory testbed, or an individual module---and must be deployed on a structurally related but physically different system with distinct dynamics, actuators, and sensors. Moreover, we bound the expected output distance between the systems using novel stochastic simulation functions.

Closest to our work is the framework in \cite{julius2009approximations}, which introduces stochastic simulation functions to quantify the output distance between partially observed stochastic hybrid systems. The setting in \cite{julius2009approximations} considers no measurement noise and treats the higher-layer system as an approximate abstraction of the lower-layer system. We propose a different notion of stochastic simulation functions tailored to systems with measurement noise and state estimators. Beyond that, we define controllers that enable the lower-layer system to remain within computable distance bounds from the higher-layer system. For the class of linear systems with Kalman  estimators, we also provide a systematic method of constructing these controllers and the respective simulation functions. 
\medskip 

\noindent{\bf Notation.} The norm $\norm{\cdot}$ is the Euclidean norm when applied to vectors and the spectral norm when applied to matrices, while $\norm{\cdot}_F$ is the Frobenius norm. $\mathbb{I}_d$ denotes the identity matrix of size $d$. $\trace(A)$ denotes the trace of a square matrix $A\in\setR^{d\times d}$. Expectation with respect to all the randomness of the underlying probability space is denoted by $\E$.

\section{Problem Formulation}

Consider the layered control architecture illustrated in Fig.~\ref{fig:problem_architecture}. The layers consist of stochastic systems $\Sigma^{(i)}$ with continuously differentiable dynamics $f^{(i)}(\cdot,\cdot)$ and locally Lipschitz continuous functions $h^{(i)}(\cdot)$. The output $y_t^{(i)}\in\setR^p$ represents the observation of the state $x_t^{(i)}\in\setR^{n_i}$, whereas $u_t^{(i)}\in\setR^{m_i}$ is the control input of $\Sigma^{(i)}$. The systems $\Sigma^{(1)}$ and $\Sigma^{(2)}$ share the same output dimension, although their states and control inputs may have different dimensions. The process noise $w_t^{(i)}\in\setR^{n_i}$ and measurement noise $v_t^{(i)}\in\setR^p$ are each independently drawn from a zero mean Gaussian distribution with covariance $\Sigma_w^{(i)}$ and $\Sigma_v^{(i)}$, respectively. We assume that the stochastic processes $\{w_t^{(1)}\}, \{v_t^{(1)}\}, \{w_t^{(2)}\},$ and $ \{v_t^{(2)}\}$ are mutually independent. The initial state $x_0^{(i)}$ is Gaussian with mean $\mu_0^{(i)}$ and covariance $\Sigma_0^{(i)}$, and is independent of all noises. 

Since the system states $x_t^{(i)}$ are not directly measured, observers can be employed to produce estimates $\hatx_t^{(i)}$ through the measurements $y_0^{(i)},\ldots,y_t^{(i)}$. A typical observer takes the  form  \cite{lesser2015controller}:
\begin{equation}\label{eq:observer}
    \hatx_{t+1}^{(i)}=f^{(i)}(\hatx_t^{(i)},u_t^{(i)})+L_t^{(i)}(y_t^{(i)}-h^{(i)}(\hatx_t^{(i)})),
\end{equation} 
where $\hatx_0^{(i)}=\mu_0^{(i)}$ and $L_t^{(i)}\in\setR^{n_i\times p}$ is the observer gain. The first term on the right-hand side of \eqref{eq:observer} is the next system state from $\hatx_t^{(i)}$ in the absence of process noise, while the second term corrects the estimation using the deviation between the actual measurement $y_t^{(i)}$ and the predicted measurement $h^{(i)}(\hatx_t^{(i)})$. Given observers of the form \eqref{eq:observer} for $\Sigma^{(i)}$, we can formalize our problem as follows.

\begin{problem}[Layered Control of Partially Observed Stochastic Systems]\label{problem}
Consider the layered control architecture depicted in Fig.~\ref{fig:problem_architecture}. Assume that observers of the form \eqref{eq:observer} are given for $\Sigma^{(1)}$ and $\Sigma^{(2)}$. Let $\{u_t^{(1)}\}$ be a sequence of inputs applied to $\Sigma^{(1)}$, with values in a compact set $\calU^{(1)} \subseteq \setR^{m_1}$. Design a controller $\pi^{(2)}:\calU^{(1)}\times\setR^{n_1}\times\setR^{n_2}\to\setR^{m_2}$ mapping triplets $(u_t^{(1)}, \hatx_t^{(1)}, \hatx_t^{(2)})$ to inputs $u_t^{(2)}$ for $\Sigma^{(2)}$ so that:
\begin{equation}\label{eq:output_guarantee}
\sup_{t\geq 0} \E\Big[\norm{y_t^{(1)} - y_t^{(2)}}\Big] \leq \varepsilon,
\end{equation}
for some $\varepsilon > 0$. The value of $\varepsilon$ may depend on $\mu_0^{(i)}$, $f^{(i)}(\cdot,\cdot)$, $h^{(i)}(\cdot)$, $\Sigma_w^{(i)}$, $\Sigma_v^{(i)}$, $L_t^{(i)}$, for $i=1,2$, and on $\calU^{(1)}$.
\end{problem}

Next, we define controllers $\pi^{(2)}(\cdot,\cdot,\cdot)$ and corresponding bounds $\varepsilon$ that address Problem~\ref{problem}. This is achieved by employing a novel notion of stochastic  Lyapunov-like functions. %\charis{Control policy or interface?}\tas{you could just mention it is also called interface in the literature}

\section{Layered Control of Partially Observed Stochastic Systems}

In this section, we present a control framework that enables  $\Sigma^{(2)}$ to approximate the output behavior of  $\Sigma^{(1)}$ within guaranteed computable  bounds (see \eqref{eq:output_guarantee}). To this end, we introduce simulation functions tailored to partially observed stochastic systems.

Intuitively, a simulation function of $\Sigma^{(1)}$ by $\Sigma^{(2)}$ is a control Lyapunov-like function defined over their state spaces. Below, we present a novel notion of stochastic simulation functions that are defined on state estimates of systems $\Sigma^{(1)}$ and $\Sigma^{(2)}$.

\begin{definition}[Stochastic Simulation Functions for Partially Observed Systems]\label{def:stochastic_simulation_function}
Consider the systems $\Sigma^{(1)}$ and $\Sigma^{(2)}$ in Fig.~\ref{fig:problem_architecture} and corresponding observers from \eqref{eq:observer}. A function $V:\setR^{n_1}\times\setR^{n_2}\to\setR_+$ is called a stochastic simulation function of $\Sigma^{(1)}$ by $\Sigma^{(2)}$ if there exists a controller $\pi^{(2)}:\calU^{(1)}\times\setR^{n_1}\times\setR^{n_2}\to\setR^{m_2}$ such that, for all input sequences $\{u_t^{(1)}\}$ taking values in $\calU^{(1)}$, the resulting stochastic processes $\{x_t^{(1)}\}$, $\{\hatx_t^{(1)}\}$, $\{x_t^{(2)}\}$, and $\{\hatx_t^{(2)}\}$ satisfy:
\begin{subequations}
\begin{align}
\label{eq:SF_condition_1}
    &\E\Big[V(\hatx_t^{(1)},\hatx_t^{(2)})\Big]\geq \E\Big[\norm{h^{(1)}(x_t^{(1)})-h^{(2)}(x_t^{(2)})}^2\Big]\\
\label{eq:SF_condition_2}
&\E\Big[V(\hatx_{t+1}^{(1)},\hatx_{t+1}^{(2)})\Big]\leq\rho \E\Big[V(\hatx_t^{(1)},\hatx_t^{(2)})\Big]+\alpha,
\end{align}
\end{subequations}
for some $\rho\in(0,1)$ and $\alpha>0$.    
\end{definition}

Condition~\eqref{eq:SF_condition_1} implies that the expectation of $V(\hatx_t^{(1)},\hatx_t^{(2)})$  bounds the mean-squared output distance between $\Sigma^{(1)}$ and $\Sigma^{(2)}$ in the absence of measurement noise. In turn, condition~\eqref{eq:SF_condition_2} guarantees that for any control sequence applied to $\Sigma^{(1)}$, the inputs generated by $\pi^{(2)}(\cdot,\cdot,\cdot)$ contract this expectation up to an additive term $\alpha$. 

By applying the controller $\pi^{(2)}(\cdot,\cdot,\cdot)$ to $\Sigma^{(2)}$, we can ensure that the expected output distance between $\Sigma^{(1)}$ and $\Sigma^{(2)}$ remains within guaranteed quantitative bounds. 

\begin{theorem}[Mean-Squared Output Tracking Guarantee]\label{thm:mean_squared_output_tracking_guarantee}
Let $V:\setR^{n_1}\times\setR^{n_2}\to\setR_+$ be a stochastic simulation function of $\Sigma^{(1)}$ by $\Sigma^{(2)}$. Moreover, let $\pi^{(2)}:\calU^{(1)}\times\setR^{n_1}\times\setR^{n_2}\to\setR^{m_2}$ be a corresponding controller for $\Sigma^{(2)}$. For any control sequence $\{u_t^{(1)}\}$ with values in $\calU^{(1)}$, the stochastic processes $\{y_t^{(1)}\}$ and $\{y_t^{(2)}\}$ satisfy:
\begin{align*}%\label{eq:output_tracking_guarantee_SF}
    \sup_{t\geq0}\E\Big[\norm{y_t^{(1)}-y_t^{(2)}}\Big]&\leq\varepsilon
\end{align*}
with:
\begin{align}\label{eq:error_bound}
   \varepsilon=\sqrt{\max\left\{V(\mu_0^{(1)},\mu_0^{(2)}),\frac{\alpha}{1-\rho}\right\}+\trace(\Sigma_v^{(1)}+\Sigma_v^{(2)})}.
\end{align} 
\end{theorem}

\begin{remark}[Result Interpretation]\label{rem:result_interpretation} Our layered control framework guarantees a uniform bound on the expected output distance between $\Sigma^{(1)}$ and $\Sigma^{(2)}$ over all $t \geq 0$. According to  \eqref{eq:error_bound}, the output behavior of $\Sigma^{(2)}$ more closely resembles the output behavior of $\Sigma^{(1)}$ whenever: i) the value of the simulation function at the expected initial states is smaller, ii) the additive term $\alpha$ in \eqref{eq:SF_condition_2} is smaller, iii) the contraction rate $1-\rho$ of ~$\E[V(\hatx_t^{(1)},\hatx_t^{(2)})]$~is larger, and iv) the trace of the measurement noise covariances is smaller.
\end{remark}

\begin{remark}
Theorem~\ref{thm:mean_squared_output_tracking_guarantee} provides a priori guarantees on the expected output distance between $\Sigma^{(1)}$ and $\Sigma^{(2)}$. In particular, if $\Sigma^{(1)}$ tracks a reference  with a maximum error of $\varepsilon' \ge 0$, the expected tracking error of $\Sigma^{(2)}$ is bounded by $\varepsilon + \varepsilon'$. The bound $\varepsilon$ can thus serve as a measure of how ``similar'' two systems are for the purpose of layering. When $\varepsilon$  is sufficiently small, the layered control design does not significantly degrade the desired output behavior. When it is not, the architecture may need to be redesigned or layering avoided altogether.
\end{remark}

The applicability of our framework relies on the ability to compute stochastic simulation functions and corresponding  controllers for $\Sigma^{(2)}$. In the following section, we provide systematic characterizations of these concepts for the class of linear systems.

\section{Stochastic Simulation Functions for Partially Observed Linear Systems}

We now specialize stochastic simulation functions to partially observed \emph{linear} systems. For this class, we provide a systematic construction of such functions along with the corresponding policies for $\Sigma^{(2)}$, yielding computable bounds on the expected output distance between $\Sigma^{(1)}$ and $\Sigma^{(2)}$.

\begin{figure}[tbh]
\begin{center}
    \begin{tikzpicture}
        % Top rectangle
        \node[rounded corners=8pt,fill=teal!20, thick, minimum width=4cm, minimum height=2cm] (top) at (0,2) {
            \begin{minipage}{7cm}
                \centering
                \vspace{-0.3cm}
                \begin{align*}
                    \hspace{.2cm}\Sigma^{(1)}:\left\{
                        \begin{array}{lcr}
                            x_{t+1}^{(1)}=A^{(1)}x_t^{(1)}+B^{(1)}u_t^{(1)}+w_t^{(1)} \\
                            y_t^{(1)} = C^{(1)}x_t^{(1)}+v_t^{(1)}
                        \end{array}
                        \right. 
                \end{align*}
            \end{minipage}
        };
        
        % Bottom rectangle
        \node[rounded corners=8pt,fill=mustard!30, thick, minimum width=4cm, minimum height=2cm] (bottom) at (0, -1.2) {
            \begin{minipage}{7.cm}
                \centering
                \vspace{-0.3cm}
                \begin{align*}
                        \hspace{.2cm}\Sigma^{(2)}:\left\{
                        \begin{array}{lcr}
                            x_{t+1}^{(2)}=A^{(2)}x_t^{(2)}+B^{(2)}u_t^{(2)}+w_t^{(2)} \\
                            y_t^{(2)} = C^{(2)}x_t^{(2)}+v_t^{(2)}
                        \end{array}
                        \right.
                \end{align*}
            \end{minipage}
        };
        
        % Shorter arrow with text
        \draw[->, thick] (top.south) -- (bottom.north) node[midway, right] {$\hatx_t^{(1)},u_t^{(1)}$};
        % Dashed arrow looping back
        %\draw[dashed, ->] (bottom.south) -- ++(0,-0.5) -- ++(4,0) -- ++(0,7.1) node[midway, left] {$\barX_+,\;\barU_+$} -- ++(-4,0) -- (top.north);
    \end{tikzpicture}
\end{center}
 \caption{Two-layer control architecture with partially observed stochastic linear systems. The Kalman state estimate $\hatx_t^{(1)}$ depends on the observed sequence $\{(y_\tau^{(1)},u_\tau^{(1)})\}_{\tau=0}^{t-1}$.  
 }
\label{fig:problem_architecture_linear}
\end{figure}
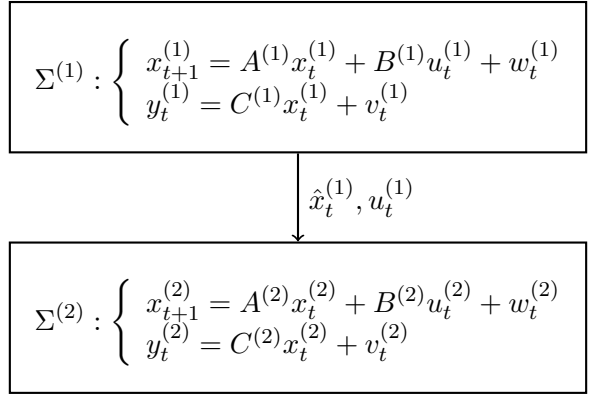

Consider the stochastic linear systems depicted in Fig.~\ref{fig:problem_architecture_linear}, where $A^{(i)}$, $B^{(i)}$, and $C^{(i)}$ are system matrices of suitable dimensions.  The observer \eqref{eq:observer} is assumed to be the well-known Kalman filter with constant gain:
\begin{equation}\label{eq:kalman_gain}
    L^{(i)} := A^{(i)}\Sigma_e^{(i)}C^{(i)\intercal}(C^{(i)}\Sigma_e^{(i)}C^{(i)\intercal}+\Sigma_v^{(i)})^{-1}
\end{equation}
in steady state, where $\Sigma_e^{(i)}$ is the  solution of the discrete algebraic Riccati equation:
\begin{align}\label{eq:kalman_riccati}
\Sigma_e^{(i)} = &(A^{(i)} - L^{(i)}C^{(i)})\Sigma_e^{(i)}(A^{(i)} - L^{(i)}C^{(i)})^{\intercal}+ \Sigma_w^{(i)} + L^{(i)}\Sigma_v^{(i)}L^{(i)\intercal}.
\end{align} 
\begin{assumption}\label{ass:kalman_ss}
 We assume that the initial state covariance is $\Sigma_0^{(i)}=\Sigma_e^{(i)}$, for $i=1, 2$.
\end{assumption}

The Kalman filter reaches steady state exponentially fast~\cite{anderson2005optimal}. Setting $\Sigma_0^{(i)} = \Sigma_e^{(i)}$ initializes the Kalman filter at steady state, yielding the constant gain \eqref{eq:kalman_gain} from $t=0$. This amounts to assuming the observer was running before the experiment, which is standard in prior work~\cite{knudsen2001consistency}.

\begin{assumption}\label{ass:relation_for_linear_systems}
System $\Sigma^{(2)}$ is stabilizable. Moreover, there exist matrices $P\in\setR^{n_2\times n_1}$ and $Q\in\setR^{m_2\times n_1}$ satisfying:
\begin{subequations}
\begin{align}
\label{eq:P_condition}
&C^{(2)}P=C^{(1)}, \\
\label{eq:P_Q_condition}
&PA^{(1)}=A^{(2)}P+B^{(2)}Q.
\end{align}
\end{subequations}
\end{assumption}

Stabilizability of $\Sigma^{(2)}$ is a minimal assumption. Conditions \eqref{eq:P_condition} and \eqref{eq:P_Q_condition} are reasonable in applications of interest (see, e.g., examples in Section~\ref{sec:Case Studies}); they are also standard in prior work \cite{Stamouli2025,lavaei2017compositional,Girard2009}. However, we note that ideas from \cite{zhong2024hierarchical} could be employed to remove condition \eqref{eq:P_Q_condition} in future work.

Before presenting our main result, we introduce the following lemma from \cite{stamouli2025arxiv}.

\begin{lemma}\label{lem:M_K_conditions} There exists a matrix $K\in\setR^{m_2\times n_2}$, a positive definite matrix $M\in\setR^{n_2\times n_2}$, and a scalar $\lambda\in(0,1)$ such that the following matrix
inequalities hold:
\begin{subequations}
\begin{align}
    \label{eq:M_condition}
    &M\succeq C^{(2)\intercal}C^{(2)}, \\
    \label{eq:M_K_condition}
    &(A^{(2)}+B^{(2)}K)^{\intercal}M(A^{(2)}+B^{(2)}K)-M\preceq-\lambda M.
\end{align}
\end{subequations}
\end{lemma}

For completeness of presentation, the proof is given in Appendix~\ref{app:lem_M_K_conditions}. The quantities $K$, $M$, and $\lambda$ can be computed based on the proof or an alternative approach presented later in this section (see Proposition~\ref{lem:optimal_M_K}). %Alternatively, an approach using semidefinite programming can be used to obtain parameters leading to a tighter bound $\varepsilon$.

Let us define the controller for system $\Sigma^{(2)}$ as follows:
\begin{equation}\label{eq:control_policy}
    \pi^{(2)}(u_t^{(1)},\hatx_t^{(1)},\hatx_t^{(2)}) = Ru_t^{(1)}+Q\hatx_t^{(1)}+K(\hatx_t^{(2)}-P\hatx_t^{(1)}),
\end{equation}
where $R\in\setR^{m_2\times m_1}$ is an arbitrary matrix and $K\in\setR^{m_2\times n_2}$ satisfies the conditions of Lemma~\ref{lem:M_K_conditions}. The first two terms drive the input of $\Sigma^{(2)}$ toward a reference point determined by $\Sigma^{(1)}$, while the last term corrects for deviations between the state estimate of $\Sigma^{(2)}$ and that of $\Sigma^{(1)}$ lifted through $P$. Let $\Sigma_\nu^{(i)}$ denote the covariance of the innovation $\nu_t^{(i)}:=y_t^{(i)}-C^{(i)}\hatx_t^{(i)}$ of the Kalman filter, that is: \begin{equation}\label{eq:innovation_cov}     \Sigma_\nu^{(i)} := C^{(i)}\Sigma_e^{(i)}C^{(i)\intercal}+\Sigma_v^{(i)}. \end{equation} The following theorem states that the controller \eqref{eq:control_policy} is associated with a stochastic simulation function, thus ensuring a computable bound $\varepsilon$ of the form \eqref{eq:error_bound}.  

\begin{theorem}[Stochastic Simulation Functions for Partially Observed Linear Systems]\label{thm:layered_control_linear}
Consider the layered control architecture in Fig.~\ref{fig:problem_architecture_linear} under Assumptions~\ref{ass:kalman_ss} and~\ref{ass:relation_for_linear_systems}. Let $K\in\setR^{m_2\times n_2}$, $M\in\setR^{n_2\times n_2}$, and $\lambda\in(0,1)$ satisfy the conditions of Lemma~\ref{lem:M_K_conditions}. Then, the function:
\begin{equation}\label{eq:simulation_function}
    V(\hatx_t^{(1)},\hatx_t^{(2)})=(\hatx_t^{(2)}-P\hatx_t^{(1)})^{\intercal}M(\hatx_t^{(2)}-P\hatx_t^{(1)})+\trace(S),
\end{equation}
where $S=\sum_{i=1}^2C^{(i)}\Sigma_e^{(i)}C^{(i)\intercal}$, is a stochastic simulation function of $\Sigma^{(1)}$ by $\Sigma^{(2)}$ with corresponding controller \eqref{eq:control_policy}, contraction scalar:
\begin{equation}\label{eq:rho}
    \rho=\frac{1-\lambda}{1-0.5\lambda},
\end{equation}
and additive scalar:
\begin{subequations}\label{eq:alpha}
\begin{align}
    \label{eq:alpha_1}
        \alpha &= \frac{2}{\lambda}\norm{M^{1/2}(B^{(2)}R-PB^{(1)})}^2(u_{\max}^{(1)})^2+\frac{\lambda}{2-\lambda}\trace(S)\\
    \label{eq:alpha_2}
    &+\norm{M^{1/2}PL^{(1)}\Sigma_\nu^{(1)1/2}}_F^2+\norm{M^{1/2}L^{(2)}\Sigma_\nu^{(2)1/2}}_F^2,
\end{align}
\end{subequations}
where $u_{\max}^{(1)}=\max_{u^{(1)}\in\,\calU^{(1)}}\norm{u^{(1)}}$.
\end{theorem}

The contraction scalar $\rho$ is determined by the stability margin of $A^{(2)}+B^{(2)}K$. Specifically, a larger stability margin allows for a larger choice of $\lambda$, resulting in a smaller value of $\rho$ and, therefore, faster contraction. The additive scalar $\alpha$ accounts for the effect of: i) the control input mismatch between $\Sigma^{(1)}$ and $\Sigma^{(2)}$, which vanishes when $B^{(2)}R = PB^{(1)}$ (see \eqref{eq:alpha_1}), ii) the output estimation errors of the Kalman filters, captured by $\trace(S)$ (see \eqref{eq:alpha_1}), and iii) the innovation covariances $\Sigma_\nu^{(i)}$ (see \eqref{eq:alpha_2}). Given matrices $M$ and $P$, the first term in $\alpha$ is minimal for:
\begin{equation}\label{eq:R}
    R=\argmin_{R'} ||M^{1/2}(B^{(2)}R'-PB^{(1)})||^2. 
\end{equation}

Theorem~\ref{thm:mean_squared_output_tracking_guarantee} guarantees that controllers of the form \eqref{eq:control_policy} maintain the expected output distance between $\Sigma^{(1)}$ and $\Sigma^{(2)}$ bounded by $\varepsilon$, for \textit{any} controller applied to system $\Sigma^{(1)}$. The bound $\varepsilon$ from \eqref{eq:error_bound} depends on the scalars $\rho$ and $\alpha$ interpreted above. The following proposition states that semidefinite programming can be used to obtain the values of $M$ and $K$ that minimize $\varepsilon$ for a given $\lambda\in(0,1)$.

\begin{proposition}\label{lem:optimal_M_K}
Fix any $\lambda\in(0,1)$. There exists a semidefinite program whose solution (if any) provides the parameters $M^{\star}$ and $K^{\star}$ that minimize the distance bound \eqref{eq:error_bound}.
\end{proposition}

For the derivation and form of this semidefinite program, see Appendix~\ref{app:Computation of the Optimal Distance Bound}. Lemma~\ref{lem:M_K_conditions} guarantees feasibility of this program for some $\lambda \in (0,1)$. An explicit feasible $\lambda$ can be obtained from the constructive argument in the proof of \Cref{lem:M_K_conditions} (see Appendix~\ref{app:lem_M_K_conditions}). The resulting semidefinite program can be solved efficiently using off-the-shelf solvers, such as the CVX toolbox \cite{diamond2016cvxpy}. To minimize $\varepsilon$ jointly over $(M,K,\lambda)$, it suffices to perform a one-dimensional search over $\lambda \in (0,1)$, solving the program for each value. 

Algorithm~\ref{alg:policy_design} summarizes the overall process for computing the control parameters and the distance bound $\varepsilon$. 

\renewcommand{\algorithmicrequire}{\textbf{Input:}}
\renewcommand{\algorithmicensure}{\textbf{Output:}}

\begin{algorithm}[H]
\caption{Layered Control Design for Partially Observed Stochastic Linear Systems}\label{alg:policy_design}
\begin{algorithmic}[1]
\Require System parameters $A^{(i)}, B^{(i)}, C^{(i)}, \mu_0^{(i)}, \Sigma_w^{(i)}, \Sigma_v^{(i)}$ for $i=1,2$; Input bound $u_{\max}^{(1)}$
\Ensure Control parameters $(P, Q, R, K)$; Bound $\varepsilon$

\Statex \textit{// Steady-state Kalman filters}
\For{$i=1,2$}
    \State Compute $\Sigma_e^{(i)}$ by solving \eqref{eq:kalman_riccati}
    \State Compute  $L^{(i)}$ from \eqref{eq:kalman_gain}
\EndFor

\Statex \textit{// Control parameters}
\State Find $P$, $Q$ satisfying \eqref{eq:P_condition}--\eqref{eq:P_Q_condition}
\State Find $K$, $M$, $\lambda$ satisfying \eqref{eq:M_condition}--\eqref{eq:M_K_condition}
\State Compute $R$ from \eqref{eq:R} 

\Statex \textit{// Simulation function parameters}
\State Compute $\rho$ from \eqref{eq:rho} and $\alpha$ from \eqref{eq:alpha}

\Statex \textit{// Error bound}
\State Compute $\varepsilon$ from \eqref{eq:error_bound}

\State \Return $(P, Q, R, K)$, $\varepsilon$
\end{algorithmic}
\end{algorithm}

In the following section, we demonstrate our method on two aerial robotic scenarios.

\section{Case Studies}\label{sec:Case Studies}
\begin{figure}[tbh]
\begin{center}
\begin{tikzpicture}
    % ===== LEFT ARCHITECTURE =====
    \begin{scope}[xshift=-3.2cm]
        % Top rectangle
        \node[rounded corners=8pt,fill=teal!20, thick, minimum width=3.2cm, minimum height=1.6cm] (topL) at (0,2) {
            \begin{minipage}{3.8cm}
                \centering
                \small $\Sigma^{(1)}$: UAV prototype
            \end{minipage}
        };
        
        % Bottom rectangle
        \node[rounded corners=8pt,fill=mustard!30, thick, minimum width=3.2cm, minimum height=1.6cm] (bottomL) at (0, -0.8) {
            \begin{minipage}{3.8cm}
                \centering
                \small $\Sigma^{(2)}$: UAV with extra control surfaces
            \end{minipage}
        };
        
        % Arrow
        \draw[->, thick] (topL.south) -- (bottomL.north);%node[midway, right] {\small $\{(y_\tau^{(1)},u_{\tau}^{(1)})\}_{\tau=0}^t$};
    \end{scope}

    % ===== RIGHT ARCHITECTURE =====
    \begin{scope}[xshift=1.7cm]
        % Top rectangle
        \node[rounded corners=8pt,fill=teal!20, thick, minimum width=3.2cm, minimum height=1.6cm] (topR) at (0,2) {
            \begin{minipage}{3.8cm}
                \centering
                \small $\Sigma^{(1)}$: Quadcopter
            \end{minipage}
        };
        
        % Bottom rectangle
        \node[rounded corners=8pt,fill=mustard!30, thick, minimum width=3.2cm, minimum height=1.6cm] (bottomR) at (0, -0.8) {
            \begin{minipage}{3.8cm}
                \centering
                %\hspace{-0.1cm}
                \small $\Sigma^{(2)}$: Hexacopter with camera payload
            \end{minipage}
        };
        
        % Arrow
        \draw[->, thick] (topR.south) -- (bottomR.north); %node[midway, right] {\small $\{(y_\tau^{(1)},u_{\tau}^{(1)})\}_{\tau=0}^t$};
    \end{scope}

    % ===== SUBFIGURE LABELS =====
    %\node at (-4.2, -2.2) {(a) First scenario};
    %\node at (4.2, -2.2) {(b) Second scenario};
\end{tikzpicture}
\end{center}
\caption{Illustration of the two aerial robot architectures.}
\label{fig:example_architectures}
\end{figure}
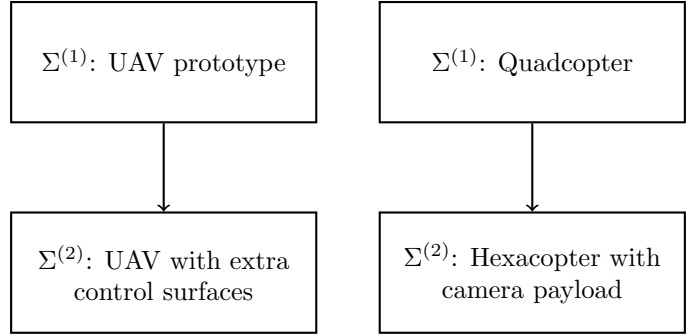

We apply our layered control framework to two aerial robots modeled as partially observed stochastic linear systems (Fig.~\ref{fig:example_architectures}). For full expressions of the models and their parameters, we refer the reader to Appendix~\ref{app:Details for Case Studies}. The code is available in our \href{https://github.com/charis-stamouli/layered-control-pos}{\textcolor{blue}{GitHub repository}}.

\begin{figure*}
    \centering
    \begin{subfigure}[t]{0.46\textwidth}
        \centering
        \includegraphics[width=\linewidth]{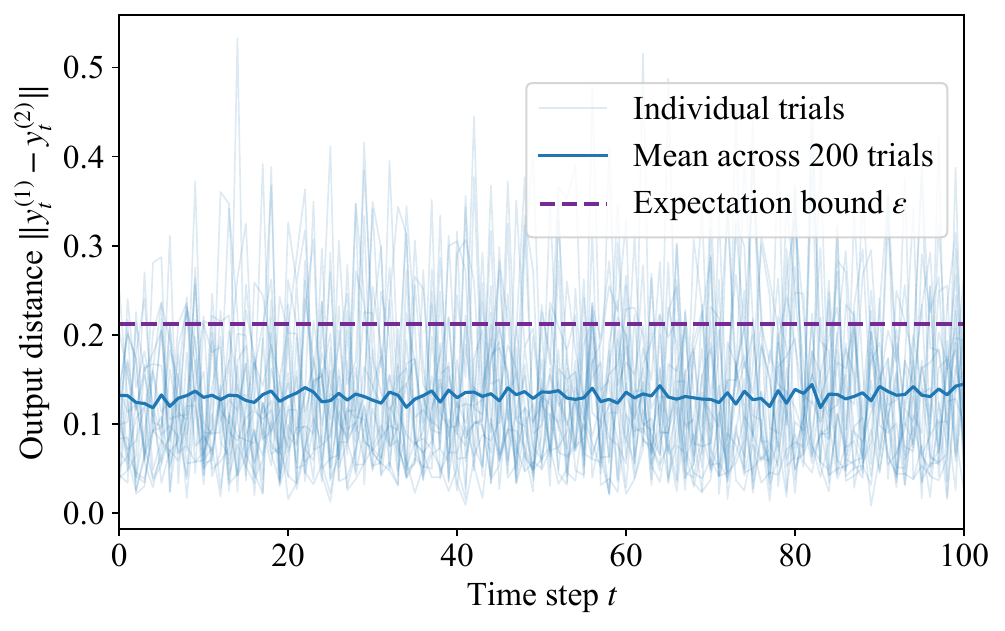}
        %\caption{Without constraint propagation.}
    \end{subfigure}
    \hspace{.5cm}
    \begin{subfigure}[t]{0.46\textwidth}
        \centering
        \includegraphics[width=\linewidth]{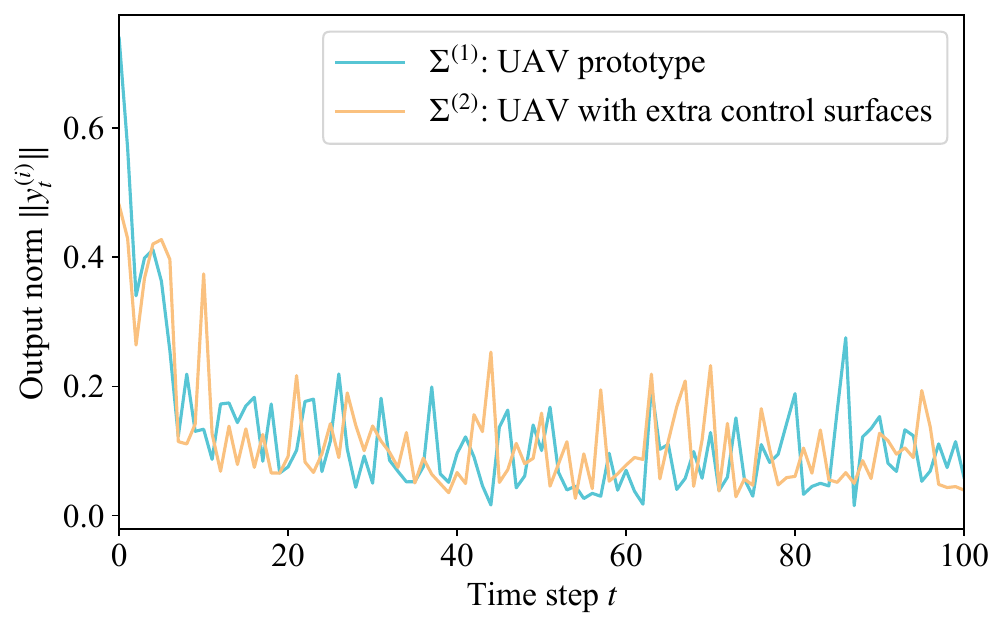}
        %\caption{With constraint propagation.}
    \end{subfigure}
    \caption{UAV with extra control surfaces. Left: Output distance 
of $\Sigma^{(1)}$ and $\Sigma^{(2)}$ for 20 individual trials, and average across 200 trials. We observe that $\varepsilon$ is a quite tight bound for the average output distance over time. 
Right: Output norm of systems $\Sigma^{(1)}$ and $\Sigma^{(2)}$ for a 
single trial. As $\Sigma^{(1)}$ is stabilized, our layered control framework leads $\Sigma^{(2)}$ to stabilize as well.}
    \label{fig:uav}
\end{figure*}

\subsection{Unmanned Aerial Vehicle (UAV)}
Consider a small fixed-wing UAV whose longitudinal dynamics are modeled with a linearized system $\Sigma^{(1)}$ from~\cite{nelson1998flight}. The state is $x^{(1)}:=(V^{(1)}, a^{(1)}, \omega_\theta^{(1)}, \theta^{(1)})$, where $V^{(1)}\in\setR$ is the airspeed, $a^{(1)}\in\setR$ the angle of attack, $\theta^{(1)}\in\setR$ the pitch angle, and $\omega_\theta^{(1)}\in\setR$ the pitch angular rate. The input $u^{(1)}:=(\delta_{\theta}, \delta_t)$ consists of the pitch actuation $\delta_{\theta}\in\setR$ and thrust actuation $\delta_t\in\setR$, constrained to the Euclidean ball of radius 4 centered at the origin. The initial state distribution has mean $\mu_0^{(1)}=(0.5, 0.08, 0.0, 0.05)$. The process and measurement noise have covariance $\Sigma_w^{(1)} := 10^{-5} \cdot \diag2(5,\, 0.5,\, 5,\, 0.5)$ and $\Sigma_v^{(1)} := 10^{-2} \cdot \diag2(10^{-2},\, 10^{-2},\, 1,\, 10^{-2})$, respectively, reflecting calm flight conditions.

We design a Linear Quadratic Gaussian (LQG) controller for $\Sigma^{(1)}$ with state penalty $P_Q = \diag2(1, 10, 5, 10)$ and input penalty $P_R = 0.1 \mathbb{I}_2$. Our goal is to leverage this controller to stabilize the same airframe after it has been upgraded with two additional control surfaces for enhanced pitch authority~\cite[Chapter 4.2]{nelson1998flight}. To this end, we model the upgraded aircraft as a linear system $\Sigma^{(2)}$ and apply our layered control framework. The state is $x^{(2)} := (V^{(2)}, a^{(2)}, \omega_\theta^{(2)}, \theta^{(2)}, \delta_{1}, \delta_{2}, \omega_{\delta1}, \omega_{\delta2})$, where the first four components are the same body states as in $\Sigma^{(1)}$, and $(\delta_{1}, \delta_{2}, \omega_{\delta1}, \omega_{\delta2})$ are the deflection angles and rates of the new surfaces. The input $u^{(2)} := (\delta_\theta, \delta_t, \tau_{1}, \tau_{2})$ extends $u^{(1)}$ with the torques $\tau_{1}, \tau_{2} \in \mathbb{R}$ driving the added surfaces. The body dynamics of $\Sigma^{(2)}$ are identical to those of $\Sigma^{(1)}$ at the nominal operating point; only the coupling between the added and original body states introduces additional dynamics. The initial state mean is $\mu_0^{(2)}=(0.5, 0.08, 0, 0.05,0,0,0,0)$. The noise covariances are $\Sigma_w^{(2)} = 10^{-5} \cdot \diag2(5,\, 0.5,\, 5,\, 0.5,\, 0.5,\, 0.5,\, 5,\, 5)$ and $\Sigma_v^{(2)} = 10^{-4} \cdot \diag2(1.5,\, 1.5,\, 100,\, 1.5)$. %The body dynamics follow the linearized longitudinal model~\cite{nelson1998flight}, and the control surface actuators are modeled as second-order torsional systems~\cite[§4.2]{nelson1998flight} coupled to the body through standard aerodynamic force and moment derivatives.

We evaluate our framework over 200 Monte Carlo trials with a horizon of $T = 100$ time steps. Theorem~\ref{thm:mean_squared_output_tracking_guarantee} yields a bound of $\varepsilon = 0.21$, indicating very similar behavior between $\Sigma^{(1)}$ and $\Sigma^{(2)}$. As shown in Fig.~\ref{fig:uav} (left), this bound is quite tight relative to the empirical mean of $\|y_t^{(1)} - y_t^{(2)}\|$ across trials. Fig.~\ref{fig:uav} (right) shows the individual output norms for a representative trial. The LQG controller stabilizes the prototype, and the controller \eqref{eq:control_policy} drives the upgraded UAV to closely track the output of the prototype, leading it to stabilize as well.
\medskip

\subsection{Hexacopter} 

\begin{figure*}[tbh]
    \centering
    \begin{subfigure}[t]{0.46\textwidth}
        \centering
        \includegraphics[width=\linewidth]{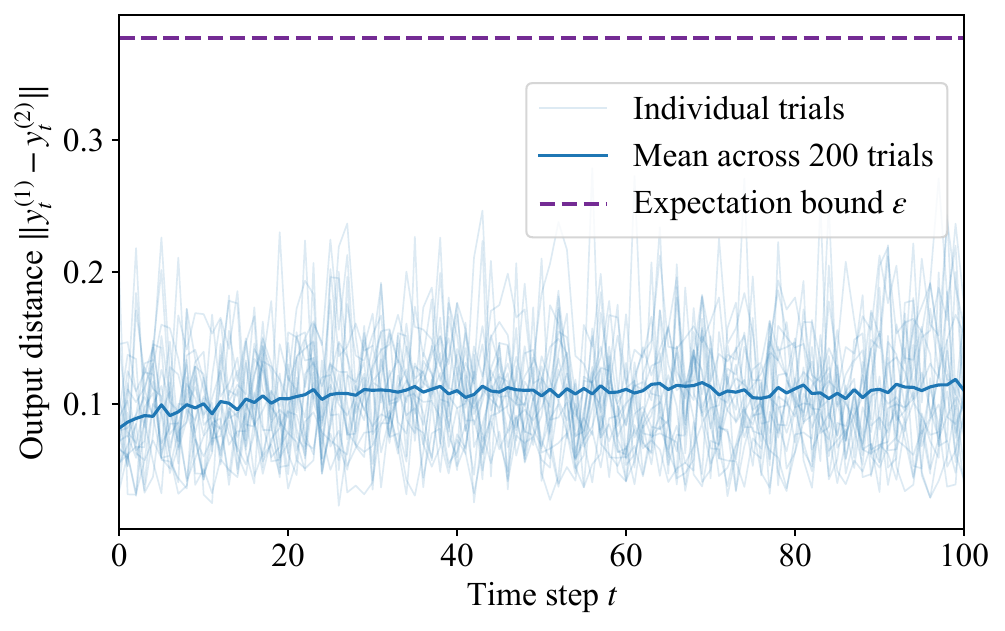}
        %\caption{Without constraint propagation.}
    \end{subfigure}
    \hspace{.5cm}
    \begin{subfigure}[t]{0.46\textwidth}
        \centering
        \includegraphics[width=\linewidth]{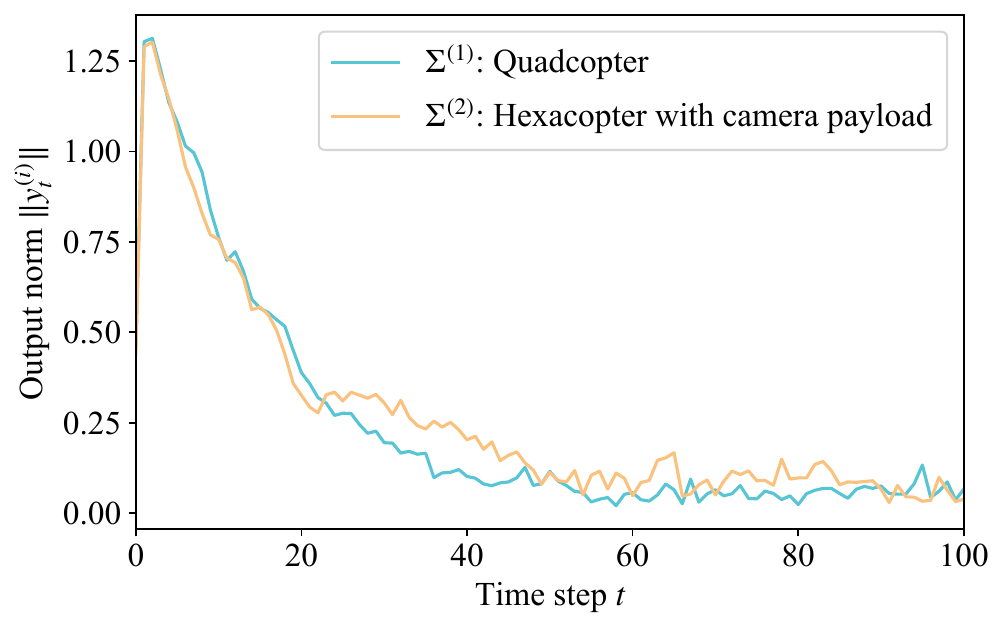}
        %\caption{With constraint propagation.}
    \end{subfigure}
    \caption{Hexacopter with camera payload. Left: Output distance 
of $\Sigma^{(1)}$ and $\Sigma^{(2)}$ for 20 individual trials, and average across 200 trials. We observe that $\varepsilon$ is a quite tight bound for the average output distance over time. 
Right: Output norm of systems $\Sigma^{(1)}$ and $\Sigma^{(2)}$ for a single trial. As $\Sigma^{(1)}$ is stabilized, our layered control framework leads $\Sigma^{(2)}$ to stabilize as well.}
    \label{fig:hexacopter}
\end{figure*}

In this scenario, the attitude and altitude dynamics of a quadcopter are 
modeled with a partially observed stochastic linear system $\Sigma^{(1)}$. 
Its state $x^{(1)} := (\phi^{(1)}, \theta^{(1)}, \omega_\phi^{(1)}, 
\omega_\theta^{(1)}, z^{(1)}, v_z^{(1)})$ consists of the roll angle 
$\phi^{(1)}$, the pitch angle $\theta^{(1)}$, their respective angular rates 
$\omega_\phi^{(1)}, \omega_\theta^{(1)}$, the altitude $z^{(1)}$, and the altitude 
velocity $v_z^{(1)}$. The 
dynamics follow from linearizing the Newton–Euler equations about 
hover~\cite{sabatino2015quadrotor,pounds2010modelling}. The 
quadcopter is actuated by the four motor thrusts contained in
$u^{(1)} := (f_1, f_2, f_3, f_4)$ with $\|u^{(1)}\| \leq 4$. The initial state distribution has mean $\mu_0^{(1)}=(0.3, 0.3, 0,0,0,0)$.
The process and measurement noise covariances are $\Sigma_w^{(1)} := 2{\cdot}10^{-6} \cdot \diag2(1,\, 1,\, 100,\, 100,\, 1,\, 100)$ and $\Sigma_v^{(1)} := 10^{-4} \cdot \diag2(1,\, 1,\, 1,\, 1,\, 25)$, respectively, reflecting calm flight conditions.

An LQG controller with state penalty $P_Q = \diag2(10, 10, 1, 1, 5, 1)$ and input penalty 
$P_R = 0.1 \mathbb{I}_4$ has been designed for $\Sigma^{(1)}$. Using our layered control framework, we aim to leverage this controller for stabilizing a hexacopter  with a  camera payload. The hexacopter is modeled as a partially 
observed stochastic linear system $\Sigma^{(2)}$, with state 
$x^{(2)} := (\phi^{(2)}, \theta^{(2)}, \omega_\phi^{(2)}, 
\omega_\theta^{(2)}, z^{(2)}, v_z^{(2)}, \alpha_1, \alpha_2, 
\omega_{\alpha_1},$ $ \omega_{\alpha_2})$, appending the camera's tilt angles 
and the angular rates \cite{nelson1998flight} to the six body states shared with $\Sigma^{(1)}$. Its input $u^{(2)} := (f_1, \ldots, f_6, \tau_1, \tau_2)$ includes the motor thrusts $f_i\in\setR$ and the camera torques 
$\tau_i\in\setR$. The initial state mean is given by $\mu_0^{(2)}=(0.3, 0.3, 0, 0,0,0,0,0,0,0)$.  The noise covariances are defined as $\Sigma_w^{(2)} := 10^{-4} \cdot \diag2(0.03,\, 0.03,\, 2.4,\, 2.4,\, 0.03,\, 2.4,\, 0.2,\, 0.2,\, 2,\, 2)$ and $\Sigma_v^{(2)} := 10^{-4} \cdot \diag2(1.5,\, 1.5,\, 1.5,\, 1.5,\, 25)$. %The hexacopter body dynamics follow the same Newton--Euler framework as the quadrotor~\cite{sabatino2015quadrotor}, with modified parameters and a six-motor mixing geometry. The gimbal is modeled as a torsional second-order system~\cite[§4.2]{nelson1998flight} whose restoring stiffness arises from the gravitational torque on the offset camera mass. The body--gimbal coupling follows from linearizing the gravity and viscous torques at the pivot.

Applying Theorem~\ref{thm:mean_squared_output_tracking_guarantee} yields 
a bound of $\varepsilon = 0.38$ on the expected output 
distance between $\Sigma^{(1)}$ and $\Sigma^{(2)}$, ensuring a priori that the systems behave similarly under our framework. We validate this bound 
empirically over 200 Monte Carlo trials with horizon $T = 100$. Fig.~\ref{fig:hexacopter} (left) shows that $\varepsilon$ closely 
upper bounds the average output distance across trials. Fig.~\ref{fig:hexacopter} (right) illustrates the output norms of each system over time for a single trial. Observe that the quadcopter is stabilized by the LQG 
controller, and the policy~\eqref{eq:control_policy} steers the hexacopter 
to reproduce the quadcopter's output, achieving stabilization in the process.

\section{Future Work}
Our work opens several directions for future research. A natural next step is to construct stochastic simulation functions for nonlinear systems, potentially leveraging reinforcement learning tools for Lyapunov function synthesis. It is also open whether the proposed layered approach can enable policy transfer across systems with unknown partially observed stochastic dynamics. Furthermore, our framework mirrors the structure of optimal transport, which seeks couplings between probability measures that minimize expected distance. Exploring this connection is a promising direction. Finally, extending the proposed framework to general partially observed Markov decision processes is an interesting avenue.

\section*{Acknowledgements}
Charis Stamouli and George J. Pappas acknowledge support from AFOSR (FA9550261B038).

\bibliographystyle{IEEEtran} % use IEEEtran.bst style
\bibliography{arxiv_version}

\appendix
\bigskip
\medskip
\noindent\textbf{\Large Appendix}
\section{Proofs}\label{app:Proofs}
\subsection{Proof of Theorem~\ref{thm:mean_squared_output_tracking_guarantee}}
From Jensen's inequality and the definition of systems $\Sigma^{(i)}$, we can write:
\begin{align}\label{thm1:proof_1}
    \E\Big[\norm{y_t^{(1)}-y_t^{(2)}}\Big]&\leq\sqrt{\E\Big[\norm{y_t^{(1)}-y_t^{(2)}}^2\Big]}=\sqrt{\E\Big[\norm{h^{(1)}(x_t^{(1)})+v_t^{(1)}-h^{(2)}(x_t^{(2)})-v_t^{(2)}}^2\Big]}.
\end{align}
The states $x_t^{(1)}$ and $x_t^{(2)}$ are independent of the measurement noises $v_t^{(1)}$ and $v_t^{(2)}$. Moreover, $v_t^{(1)}$ is independent of $v_t^{(2)}$. Therefore, \eqref{thm1:proof_1} yields:
\begin{align}\label{thm1:proof_2}
    \E\Big[\norm{y_t^{(1)}-y_t^{(2)}}\Big]&\leq\sqrt{\E\Big[\norm{h^{(1)}(x_t^{(1)})-h^{(2)}(x_t^{(2)})}^2+\norm{v_t^{(1)}}^2+\norm{v_t^{(2)}}^2\Big]}\nonumber\\
    &= \sqrt{\E\Big[\norm{h^{(1)}(x_t^{(1)})-h^{(2)}(x_t^{(2)})}^2\Big]+\trace(\Sigma_v^{(1)}+\Sigma_v^{(2)})}\nonumber\\
    &\leq\sqrt{ \E[V(\hatx_t^{(1)},\hatx_t^{(2)})]+\trace(\Sigma_v^{(1)}+\Sigma_v^{(2)})},
\end{align}
where the last inequality follows from \eqref{eq:SF_condition_1}. By expanding \eqref{eq:SF_condition_2} recursively, we obtain:
\begin{align}\label{thm1:proof_3}
\E[V(\hatx_t^{(1)},\hatx_t^{(2)})]&\leq\rho^t\E[V(\hatx_0^{(1)},\hatx_0^{(2)})]+\alpha\sum_{\tau=0}^{t-1}\rho^{\tau}=\rho^t\E[V(\hatx_0^{(1)},\hatx_0^{(2)})]+\frac{\alpha(1-\rho^t)}{1-\rho}\nonumber\\
    &=\rho^tV(\mu_0^{(1)},\mu_0^{(2)})+\frac{\alpha(1-\rho^t)}{1-\rho},
\end{align}
given that $\hatx_0^{(i)}=\mu_0^{(i)}$, for $i=1,2$. Since $\rho\in(0,1)$, we have $\rho^t\in(0,1)$, and thus the right-hand side of \eqref{thm1:proof_3} is a convex combination of $V(\mu_0^{(1)},\mu_0^{(2)})$ and $\alpha/(1-\rho)$. Hence, we have:
\begin{equation*}
    \E[V(\hatx_t^{(1)},\hatx_t^{(2)})]\leq\max\left\{V(\mu_0^{(1)},\mu_0^{(2)}),\frac{\alpha}{1-\rho}\right\},
\end{equation*}
which results in \eqref{thm1:proof_2} giving:
\begin{equation}
    \E\Big[\norm{y_t^{(1)}-y_t^{(2)}}\Big]\leq\sqrt{ \max\left\{V(\mu_0^{(1)},\mu_0^{(2)}),\frac{\alpha}{1-\rho}\right\}+\trace(\Sigma_v^{(1)}+\Sigma_v^{(2)})},
\end{equation}
thus completing the proof of the theorem.

\subsection{Proof of Lemma~\ref{lem:M_K_conditions}}\label{app:lem_M_K_conditions}
The proof follows the same steps as the proof of \cite[Lemma 2]{stamouli2025arxiv}. By Assumption~\ref{ass:relation_for_linear_systems}, system $\Sigma^{(2)}$ is stabilizable, which suggests the existence of a gain matrix $K\in\setR^{m\times n}$ such that all the eigenvalues of $A_K^{(2)}:=A^{(2)}+B^{(2)}K$ are strictly inside the unit disk. Let $M=N+C^{(2)\intercal}C^{(2)}$, where $N\in\setR^{n_2\times n_2}$ is a positive definite matrix to be determined later on. Since $N$ is positive definite, it follows that $M$ is 
positive definite and that 
inequality~\eqref{eq:M_condition} is satisfied. Define:
\[
    A_{K,\lambda}^{(2)} := \frac{1}{\sqrt{1-\lambda}} A_K^{(2)},
\]
where $\lambda\in(0,1)$ is chosen such that the eigenvalues 
of $A_{K,\lambda}^{(2)}$ lie strictly inside the unit disk. Such a scalar $\lambda$ exists since it suffices to take it 
sufficiently small. By the definitions of $M$ and 
$A_{K,\lambda}^{(2)}$, inequality~\eqref{eq:M_K_condition} 
is equivalent to:
\begin{align}
    &A_K^{(2)\intercal}MA_K^{(2)}-(1-\lambda)M\preceq0 \nonumber\\
    \iff &A_{K,\lambda}^{(2)\intercal}MA_{K,\lambda}^{(2)}-M\preceq0\nonumber\\
    \label{lem:tracking_precision_guarantee_proof_2}
    \iff &A_{K,\lambda}^{(2)\intercal}NA_{K,\lambda}^{(2)}-N\preceq-A_{K,\lambda}^{(2)\intercal}C^{(2)\intercal}C^{(2)}A_{K,\lambda}^{(2)}+C^{(2)\intercal}C^{(2)}.
\end{align}
Let $\Lambda\succeq0$ satisfy:
\begin{equation}\label{lem:tracking_precision_guarantee_proof_3}
    A_{K,\lambda}^{(2)\intercal}C^{(2)\intercal}C^{(2)}A_{K,\lambda}^{(2)}-C^{(2)\intercal}C^{(2)}\preceq \Lambda,
\end{equation}
and define $N$ as the unique positive definite solution of 
the Lyapunov equation:
\begin{equation}\label{lem:tracking_precision_guarantee_proof_4}
    A_{K,\lambda}^{(2)\intercal}NA_{K,\lambda}^{(2)}-N=-\Lambda.
\end{equation}
Substituting~\eqref{lem:tracking_precision_guarantee_proof_4} 
into~\eqref{lem:tracking_precision_guarantee_proof_3} yields 
inequality~\eqref{lem:tracking_precision_guarantee_proof_2}. 
Since \eqref{lem:tracking_precision_guarantee_proof_2} is 
equivalent to~\eqref{eq:M_K_condition}, this completes the 
proof of the lemma.

\subsection{Proof of Theorem~\ref{thm:layered_control_linear}}
In steady state, the state estimate dynamics of the Kalman filter take the form:
\begin{equation}\label{eq:kalman_observer}
    \hatx_{t+1}^{(i)}=A^{(i)}\hatx_t^{(i)}+B^{(i)}u_t^{(i)}+L^{(i)}\nu_t^{(i)},
\end{equation}
where $L^{(i)}$ is given by \eqref{eq:kalman_gain} and $\nu_t^{(i)}:=y_t^{(i)}-C^{(i)}\hatx_t^{(i)}$ is the innovation. Recall that the state estimation error of $\Sigma^{(i)}$ is defined as $e_t^{(i)}=x_t^{(i)}-\hatx_t^{(i)}$. Consider applying the controller \eqref{eq:control_policy} to system $\Sigma^{(2)}$. Define the associated gap $\Delta\hatx_t:=\hatx_t^{(2)}-P\hatx_t^{(1)}$ between the state estimate of $\Sigma^{(2)}$ and the state estimate of $\Sigma^{(1)}$ lifted through $P$. By performing straightforward algebraic manipulations, the dynamics of $\Delta\hatx_t$ can be written as follows:
\begin{align}
    \label{thm2:proof_1}
    &\Delta\hatx_{t+1} = \hatx_{t+1}^{(2)}-P\hatx_{t+1}^{(1)}\nonumber\\
    &=A^{(2)}\hatx_t^{(2)}+B^{(2)}(Ru_t^{(1)}+Q\hatx_t^{(1)}+K\Delta\hatx_t)+L^{(2)}\nu_t^{(2)}-P(A^{(1)}\hatx_t^{(1)}+B^{(1)}u_t^{(1)}+L^{(1)}\nu_t^{(1)})\nonumber\\
    &=(A^{(2)}+B^{(2)}K)\Delta\hatx_t+(A^{(2)}P+B^{(2)}Q-PA^{(1)})\hatx_t^{(1)}+(B^{(2)}R-PB^{(1)})u_t^{(1)}+L^{(2)}\nu_t^{(2)}-PL^{(1)}\nu_t^{(1)}\nonumber\\
    &=\Acl\Delta\hatx_t+(B^{(2)}R-PB^{(1)})u_t^{(1)}+L^{(2)}\nu_t^{(2)}-PL^{(1)}\nu_t^{(1)},
\end{align}
where $\Acl:=A^{(2)}+B^{(2)}K$ and the third step follows from \eqref{eq:P_Q_condition}. The gain $K$ renders $\Acl$ Schur stable, while the remaining terms in \eqref{thm2:proof_1} are determined by: i) the input mismatch $B^{(2)}R-PB^{(1)}$, and ii) the innovations $\nu_t^{(i)}$ of the two Kalman filters.
 
Throughout the proof, we fix an arbitrary sequence of control inputs $u_0^{(1)},\ldots,u_t^{(1)},\ldots$ taking values in $\calU^{(1)}$. We summarize a few remarks that will be used repeatedly:
\begin{enumerate}[label=(\roman*)]
    \item \label{remark-i} the estimation error satisfies $e_{t+1}^{(i)}=(A^{(i)}-L^{(i)}C^{(i)})e_t^{(i)}+w_t^{(i)}-L^{(i)}v_t^{(i)}$, thus not depending on the control input; hence $e_t^{(1)}$ and $\nu_t^{(1)}$ are independent of $e_t^{(2)}$ and $\nu_t^{(2)}$, since the noise processes driving $\Sigma^{(1)}$ and $\Sigma^{(2)}$ are mutually independent;
    \item \label{remark-ii} the estimation error $e_t^{(i)}$ is zero mean with covariance $\Sigma_e^{(i)}$, as given in \eqref{eq:kalman_riccati};
    \item \label{remark-iii} the estimation error $e_t^{(i)}$ is independent of the state estimates $\hatx_t^{(1)}$ and $\hatx_t^{(2)}$, as they are uncorrelated and jointly Gaussian;
    \item \label{remark-iv} the innovation satisfies:
    \begin{equation}\label{thm2:innovation_decomp}
        \nu_t^{(i)}=y_t^{(i)}-C^{(i)}\hatx_t^{(i)}=C^{(i)}x_t^{(i)}+v_t^{(i)}-C^{(i)}\hatx_t^{(i)}=C^{(i)}e_t^{(i)}+v_t^{(i)},
    \end{equation}
    and is therefore zero mean with covariance $\Sigma_\nu^{(i)}$ as defined in \eqref{eq:innovation_cov}, by \ref{remark-ii} and the independence of $e_t^{(i)}$ and $v_t^{(i)}$;
    \item \label{remark-v} the innovation $\nu_t^{(i)}$ is independent of $\Delta\hatx_t$, since $\nu_t^{(i)}$ is uncorrelated with the past observations of $\Sigma^{(i)}$ \cite{anderson2005optimal} and, by \ref{remark-i}, with those of the other system.
\end{enumerate}
 
\textbf{Condition \eqref{eq:SF_condition_1}.} We first show that the function \eqref{eq:simulation_function} satisfies condition \eqref{eq:SF_condition_1}. Using $x_t^{(i)}=\hatx_t^{(i)}+e_t^{(i)}$, we can write:
\begin{align}\label{thm2:proof_3}
    &\E\Big[\norm{C^{(1)}x_t^{(1)}-C^{(2)}x_t^{(2)}}^2\Big] = \E\Big[\norm{C^{(1)}(\hatx_t^{(1)}+e_t^{(1)})-C^{(2)}(\hatx_t^{(2)}+e_t^{(2)})}^2\Big] \nonumber\\
    &= \E\Big[\norm{C^{(1)}\hatx_t^{(1)}-C^{(2)}\hatx_t^{(2)}}^2+\norm{C^{(1)}e_t^{(1)}}^2+\norm{C^{(2)}e_t^{(2)}}^2\Big] \hspace{2cm}(\text{from \ref{remark-i} and \ref{remark-iii}}) \nonumber\\
    &= \E\Big[\norm{C^{(2)}(\hatx_t^{(2)}-P\hatx_t^{(1)})}^2\Big]+\sum_{i=1}^2\E\Big[e_t^{(i)\intercal}C^{(i)\intercal}C^{(i)}e_t^{(i)}\Big]  \hspace{2.1cm}(\text{from \eqref{eq:P_condition}}) \nonumber\\
    &=\E\Big[\Delta\hatx_t^{\intercal}C^{(2)\intercal}C^{(2)}\Delta\hatx_t\Big]+\trace(S),
\end{align}
where the last step follows from \ref{remark-ii} and the fact that:
\begin{align*}
    \sum_{i=1}^2\E\Big[e_t^{(i)\intercal}C^{(i)\intercal}C^{(i)}e_t^{(i)}\Big]&=\sum_{i=1}^2\trace\big(C^{(i)\intercal}C^{(i)}\E[e_t^{(i)}e_t^{(i)\intercal}]\big)\\
    &=\sum_{i=1}^2\trace\big(C^{(i)}\Sigma_e^{(i)}C^{(i)\intercal}\big)=\trace(S).
\end{align*}
Employing \eqref{eq:M_condition} and the definition of $V(\cdot,\cdot)$ in \eqref{eq:simulation_function}, \eqref{thm2:proof_3} leads to:
\begin{align*}
    \E\Big[\norm{C^{(1)}x_t^{(1)}-C^{(2)}x_t^{(2)}}^2\Big]
    \leq\E\Big[\Delta\hatx_t^{\intercal}M\Delta\hatx_t\Big]+\trace(S)=\E\Big[V(\hatx_t^{(1)},\hatx_t^{(2)})\Big],
\end{align*}
which completes the proof of \eqref{eq:SF_condition_1}.
 
\textbf{Condition \eqref{eq:SF_condition_2}.} Next, we show that the function \eqref{eq:simulation_function} satisfies condition \eqref{eq:SF_condition_2}. Setting:
\begin{equation}\label{eq:d_t}
    d_t := \Acl\Delta\hatx_t+(B^{(2)}R-PB^{(1)})u_t^{(1)},
\end{equation}
the dynamics \eqref{thm2:proof_1} read $\Delta\hatx_{t+1}=d_t+L^{(2)}\nu_t^{(2)}-PL^{(1)}\nu_t^{(1)}$. By \ref{remark-i}, \ref{remark-iv}, and \ref{remark-v}, the innovations $\nu_t^{(1)}$ and $\nu_t^{(2)}$ are zero mean, independent of each other, and independent of $d_t$. Hence, all cross terms in the expansion of the quadratic form below have zero mean, and we obtain:
\begin{align}
    \label{thm2:proof_5}
    &\E\Big[V(\hatx_{t+1}^{(1)},\hatx_{t+1}^{(2)})\Big]=\E\Big[\Delta\hatx_{t+1}^{\intercal}M\Delta\hatx_{t+1}\Big]+\trace(S)\nonumber\\
    &=\E\Big[d_t^{\intercal}Md_t\Big]+\E\Big[\nu_t^{(1)\intercal}L^{(1)\intercal}P^{\intercal}MPL^{(1)}\nu_t^{(1)}\Big]+\E\Big[\nu_t^{(2)\intercal}L^{(2)\intercal}ML^{(2)}\nu_t^{(2)}\Big]+\trace(S)\nonumber\\
    &=\E\Big[d_t^{\intercal}Md_t\Big]+\trace\big(L^{(1)\intercal}P^{\intercal}MPL^{(1)}\Sigma_\nu^{(1)}\big)+\trace\big(L^{(2)\intercal}ML^{(2)}\Sigma_\nu^{(2)}\big)+\trace(S)\nonumber\\
    &=\E\Big[d_t^{\intercal}Md_t\Big]+\norm{M^{1/2}PL^{(1)}\Sigma_\nu^{(1)1/2}}_F^2+\norm{M^{1/2}L^{(2)}\Sigma_\nu^{(2)1/2}}_F^2+\trace(S),
\end{align}
where the third step uses \ref{remark-iv}.
 
It remains to bound $\E[d_t^{\intercal}Md_t]$. By performing straightforward algebraic manipulations on \eqref{eq:d_t}, we get:
\begin{align}
    &d_t^{\intercal}Md_t= \Delta\hatx_t^{\intercal}A_K^{(2)\intercal}M\Acl\Delta\hatx_t+\norm{M^{1/2}(B^{(2)}R-PB^{(1)})u_t^{(1)}}^2\nonumber\\
    &\hspace{1.7cm}+2\Delta\hatx_t^{\intercal}A_K^{(2)\intercal}M(B^{(2)}R-PB^{(1)})u_t^{(1)}\nonumber\\
    &\leq \Delta\hatx_t^{\intercal}A_K^{(2)\intercal}M\Acl\Delta\hatx_t+\norm{M^{1/2}(B^{(2)}R-PB^{(1)})u_t^{(1)}}^2\nonumber\\
    \label{thm2:proof_6}
    &+2\norm{M^{1/2}\Acl\Delta\hatx_t}\norm{M^{1/2}(B^{(2)}R-PB^{(1)})u_t^{(1)}},
\end{align}
where the last step follows from a direct application of the Cauchy--Schwarz inequality. By setting:
\[
a=\frac{1}{\sqrt{\mu}}\norm{M^{1/2}\Acl\Delta\hatx_t},\quad b=\sqrt{\mu}\norm{M^{1/2}(B^{(2)}R-PB^{(1)})u_t^{(1)}},
\]
where $\mu=(2-\lambda)/\lambda$, the last term in \eqref{thm2:proof_6} can be bounded as follows:
\begin{align}
    &2\norm{M^{1/2}\Acl\Delta\hatx_t}\norm{M^{1/2}(B^{(2)}R-PB^{(1)})u_t^{(1)}}=2ab\nonumber\\
    \leq\; &a^2+b^2 \hspace{6.5cm}(\text{by Young's inequality}) \nonumber\\
    \label{thm2:proof_7}
    =\; &\frac{1}{\mu}\Delta\hatx_t^{\intercal}A_K^{(2)\intercal} M\Acl\Delta\hatx_t+\mu\norm{M^{1/2}(B^{(2)}R-PB^{(1)})u_t^{(1)}}^2.
\end{align}
Employing \eqref{thm2:proof_7} and the identities $1+1/\mu=1/(1-0.5\lambda)$ and $1+\mu=2/\lambda$, inequality \eqref{thm2:proof_6} yields:
\begin{align}
    \label{thm2:proof_8}
    d_t^{\intercal}Md_t&\leq\left(\frac{1}{1-0.5\lambda}\right)\Delta\hatx_t^{\intercal}A_K^{(2)\intercal}M\Acl\Delta\hatx_t+\frac{2}{\lambda}\norm{M^{1/2}(B^{(2)}R-PB^{(1)})u_t^{(1)}}^2.
\end{align}
Given that $\lambda\in(0,1)$, notice that \eqref{eq:M_K_condition} implies that:
\begin{align}
    &A_K^{(2)\intercal}M\Acl-M\preceq-\lambda M \nonumber\\
    \iff\; &A_K^{(2)\intercal}M\Acl-(1-0.5\lambda)M\preceq-0.5\lambda M \nonumber\\
     \iff\; &\left(\frac{1}{1-0.5\lambda}\right)A_K^{(2)\intercal}M\Acl-M\preceq-\frac{\lambda}{2-\lambda}M\nonumber\\
    \label{thm2:proof_9}
     \iff\; &\left(\frac{1}{1-0.5\lambda}\right)A_K^{(2)\intercal}M\Acl\preceq\frac{2(1-\lambda)}{2-\lambda}M=\rho M,
\end{align}
where the last equality follows from the definition \eqref{eq:rho} of $\rho$. Since $u_t^{(1)}\in\calU^{(1)}$, we have $\norm{u_t^{(1)}}\leq u_{\max}^{(1)}$, and thus the Cauchy--Schwarz inequality together with \eqref{thm2:proof_9} allows us to bound \eqref{thm2:proof_8} as:
\begin{align}
    \label{thm2:proof_9b}
    d_t^{\intercal}Md_t\leq\rho\,\Delta\hatx_t^{\intercal}M\Delta\hatx_t+\frac{2}{\lambda}\norm{M^{1/2}(B^{(2)}R-PB^{(1)})}^2(u_{\max}^{(1)})^2.
\end{align}
By definition of $V(\cdot,\cdot)$, we have $\E[\Delta\hatx_t^{\intercal}M\Delta\hatx_t]=\E[V(\hatx_t^{(1)},\hatx_t^{(2)})]-\trace(S)$. Substituting \eqref{thm2:proof_9b} into \eqref{thm2:proof_5} therefore gives:
\begin{align}
    \label{thm2:proof_10}
    &\E\Big[V(\hatx_{t+1}^{(1)},\hatx_{t+1}^{(2)})\Big]\leq\rho\,\E\Big[V(\hatx_t^{(1)},\hatx_t^{(2)})\Big]+\frac{2}{\lambda}\norm{M^{1/2}(B^{(2)}R-PB^{(1)})}^2(u_{\max}^{(1)})^2\nonumber\\
    &\quad+(1-\rho)\trace(S)+\norm{M^{1/2}PL^{(1)}\Sigma_\nu^{(1)1/2}}_F^2+\norm{M^{1/2}L^{(2)}\Sigma_\nu^{(2)1/2}}_F^2.
\end{align}
Finally, from the definition \eqref{eq:rho} of $\rho$ we obtain:
\begin{equation}\label{eq:one_minus_rho}
    1-\rho=1-\frac{2(1-\lambda)}{2-\lambda}=\frac{(2-\lambda)-2(1-\lambda)}{2-\lambda}=\frac{\lambda}{2-\lambda},
\end{equation}
so that the right-hand side of \eqref{thm2:proof_10} equals $\rho\,\E[V(\hatx_t^{(1)},\hatx_t^{(2)})]+\alpha$, with $\alpha$ given by \eqref{eq:alpha}. Consequently, condition \eqref{eq:SF_condition_2} of \Cref{def:stochastic_simulation_function} is also satisfied, which completes the proof of the theorem.

\subsection{Proof of Proposition~\ref{lem:optimal_M_K}}
For the proof, please see Appendix~\ref{app:Computation of the Optimal Distance Bound}.

\section{Computation of optimal parameters $M$ and $K$}\label{app:Computation of the Optimal Distance Bound}
 
In this section, we describe how to compute the parameters $M$ and $K$ that minimize the output distance bound \eqref{eq:error_bound} for a fixed $\lambda\in(0,1)$. Using \eqref{eq:one_minus_rho}, we deduce that this bound takes the form:
\begin{equation}\label{eq:eps_recap}
    \varepsilon \;=\; \sqrt{\max\!\left\{\,V(\mu_0^{(1)},\mu_0^{(2)}),\;\frac{(2-\lambda)\,\alpha}{\lambda}\,\right\} \;+\; \trace\!\left(\Sigma_v^{(1)} + \Sigma_v^{(2)}\right)},
\end{equation}
where $V(\mu_0^{(1)},\mu_0^{(2)})$ collects the contribution of the initial conditions and $\alpha$ collects the contributions of the inputs and the innovations (see \eqref{eq:simulation_function} and \eqref{eq:alpha}). Both $V(\mu_0^{(1)},\mu_0^{(2)})$ and $\alpha$ depend on $M$, while the contraction-type constraint \eqref{eq:M_K_condition} couples $M$ and $K$ in a bilinear way. For simplicity, we focus on minimizing $\varepsilon^2$---which is equivalent to minimizing $\varepsilon$---through the following four-step procedure using convex optimization tools from \cite{boyd2004convex}.
 
\paragraph{Step 1: Change of variables.}
We introduce:
\begin{equation}
    \tilde{M} \;=\; M^{-1}, \qquad \tilde{K} \;=\; K\,\tilde{M},
\end{equation}
which linearizes the bilinear coupling between $M$ and $K$ in \eqref{eq:M_K_condition}. A direct application of the Schur complement shows that the original constraints \eqref{eq:M_condition} and \eqref{eq:M_K_condition} are equivalent to the linear matrix inequalities (LMIs):
\begin{subequations}
\begin{align}
    \label{eq:LMI_C}
    &\begin{bmatrix}
        I & C^{(2)}\tilde{M} \\
        \tilde{M}C^{(2)\intercal} & \tilde{M}
    \end{bmatrix}\succeq0,\\[4pt]
    \label{eq:LMI_contraction}
    &\begin{bmatrix}
        \tilde{M} & A^{(2)}\tilde{M}+B^{(2)}\tilde{K}\\
        (A^{(2)}\tilde{M}+B^{(2)}\tilde{K})^{\intercal} & (1-\lambda)\tilde{M}
    \end{bmatrix}\succeq0,
\end{align}
\end{subequations}
respectively.
 
\paragraph{Step 2: Epigraph form for $\varepsilon$.}
Since \eqref{eq:eps_recap} contains a maximum of two terms, we introduce an epigraph variable $\gamma\in\mathbb{R}$ and replace the original objective by:
\begin{equation}\label{eq:epigraph_form}
    \min\;\gamma \quad\text{s.t.}\quad \gamma \;\geq\; V(\mu_0^{(1)},\mu_0^{(2)}),\qquad \gamma \;\geq\; \frac{(2-\lambda)\,\alpha}{\lambda}.
\end{equation}
The additive constant $\trace(\Sigma_v^{(1)}+\Sigma_v^{(2)})$ does not affect the minimizer and is dropped from the optimization.
 
\paragraph{Step 3: LMI reformulation of constraints from \eqref{eq:epigraph_form}.}
Setting $z_0 = \mu_0^{(2)} - P\mu_0^{(1)}$, \eqref{eq:simulation_function} implies that $V(\mu_0^{(1)},\mu_0^{(2)}) = z_0^{\intercal} M z_0 + \trace(S)$. Using $M = \tilde{M}^{-1}$ and applying the Schur complement, the inequality $\gamma \geq z_0^{\intercal} M z_0 + \trace(S)$ becomes the LMI:
\begin{equation}\label{eq:LMI_V}
    \begin{bmatrix}
        \gamma - \trace(S) & z_0^{\intercal} \\
        z_0 & \tilde{M}
    \end{bmatrix} \;\succeq\; 0.
\end{equation}
From \eqref{eq:alpha}, the quantity $\alpha$ is a linear combination of terms of the form $\bigl\|M^{1/2} X_j\bigr\|_{F}^{2} = \trace(X_j^{\intercal} M X_j)$, plus the term $\tfrac{\lambda}{2-\lambda}\trace(S)$, which does not depend on the decision variables and can be precomputed. The relevant matrices $X_j$ are:
\begin{equation}\label{eq:Xj_list}
    X_0 := B^{(2)}R - PB^{(1)},\qquad X_1 := P L^{(1)}\Sigma_\nu^{(1)\,1/2},\qquad X_2 := L^{(2)}\Sigma_\nu^{(2)\,1/2},
\end{equation}
with $\Sigma_\nu^{(i)}$ given by \eqref{eq:innovation_cov}. For each $X_j$, we introduce an auxiliary symmetric matrix $T_j \succeq 0$ together with the constraint $T_j \succeq X_j^{\intercal} M X_j$. Substituting $M = \tilde{M}^{-1}$ and applying the Schur complement once more, this constraint is equivalent to the LMI:
\begin{equation}\label{eq:LMI_Tj}
    \begin{bmatrix}
        T_j & X_j^{\intercal} \\
        X_j & \tilde{M}
    \end{bmatrix} \;\succeq\; 0.
\end{equation}
With these auxiliary variables in place, $\bigl\|M^{1/2}X_j\bigr\|_F^{2} \leq \trace(T_j)$, and the upper bound on $\alpha$ becomes an affine function of the $T_j$'s, given by:
\begin{equation}\label{eq:alpha_upper}
    \bar{\alpha} \;:=\; \frac{2\,(u_{\max}^{(1)})^{2}}{\lambda}\,\trace(T_0) \;+\; \frac{\lambda}{2-\lambda}\,\trace(S) \;+\; \trace(T_{1}) \;+\; \trace(T_{2}).
\end{equation}
Note that $\bar{\alpha}=\alpha$ at any optimal solution, since each constraint $T_j\succeq X_j^{\intercal}MX_j$ is tight at the minimum of $\trace(T_j)$.
 
\paragraph{Step 4: The final semidefinite program.}
Combining Steps 1--3, the problem of minimizing $\varepsilon$ for a fixed $\lambda\in(0,1)$ reduces to the following semidefinite program:
\begin{subequations}\label{eq:final_SDP}
\begin{align}
    \min_{\tilde{M},\,\tilde{K},\,\gamma,\,T_0,\,T_{1},\,T_{2}} \quad & \gamma \\
    \mathrm{s.t.}\quad
        & \tilde{M} \;\succeq\; \epsilon \mathbb{I}_{n_2}, \\[2pt]
        \label{eq:final_LMI_C}
        & \begin{bmatrix} I & C^{(2)}\tilde{M} \\ \tilde{M}C^{(2)\intercal} & \tilde{M} \end{bmatrix}\succeq 0, \\[2pt]
        \label{eq:final_LMI_contraction}
        & \begin{bmatrix} \tilde{M} & A^{(2)}\tilde{M}+B^{(2)}\tilde{K} \\ (A^{(2)}\tilde{M}+B^{(2)}\tilde{K})^{\intercal} & (1-\lambda)\tilde{M} \end{bmatrix}\succeq 0, \\[2pt]
        \label{eq:final_LMI_V}
        & \begin{bmatrix} \gamma - \trace(S) & z_0^{\intercal} \\ z_0 & \tilde{M} \end{bmatrix}\succeq 0, \\[2pt]
        \label{eq:final_LMI_Tj}
        & \begin{bmatrix} T_j & X_j^{\intercal} \\ X_j & \tilde{M} \end{bmatrix}\succeq 0, \quad j\in\{0,1,2\}, \\[2pt]
        \label{eq:final_alpha_branch}
        & \gamma \;\geq\; \frac{2-\lambda}{\lambda}\,\bar{\alpha},
\end{align}
\end{subequations}
where $\epsilon$ is a small positive constant and $\bar{\alpha}$ is the affine expression defined in \eqref{eq:alpha_upper}. Constraint \eqref{eq:final_LMI_V} enforces $\gamma \geq V(\mu_0^{(1)},\mu_0^{(2)})$, while constraint \eqref{eq:final_alpha_branch} enforces $\gamma \geq (2-\lambda)\alpha/\lambda$; together they realize the maximum in \eqref{eq:eps_recap}. Let $(\tilde{M}^{\star},\tilde{K}^{\star},\gamma^{\star})$ denote an optimal solution of \eqref{eq:final_SDP}. The optimal parameters of interest are then recovered as:
\begin{equation*}
    M^{\star} \;=\; (\tilde{M}^{\star})^{-1}, \qquad K^{\star} \;=\; \tilde{K}^{\star}\,(\tilde{M}^{\star})^{-1},
\end{equation*}
and the corresponding optimal value of the bound is $\varepsilon^{\star} = \sqrt{\,\gamma^{\star} + \trace(\Sigma_v^{(1)}+\Sigma_v^{(2)})\,}$.

%Feasibility of \eqref{eq:final_SDP} for some $\lambda\in(0,1)$ follows from \Cref{lem:M_K_conditions}. An explicit feasible $\lambda$ can be obtained by following the constructive argument in the proof of \Cref{lem:M_K_conditions} (see \cite[Appendix~A.2]{stamouli2025arxiv}). Since \eqref{eq:final_SDP} is a semidefinite program, it can be solved efficiently with off-the-shelf solvers, such as the CVX toolbox \cite{diamond2016cvxpy}. To minimize $\varepsilon$ jointly over $(M,K,\lambda)$, it suffices to perform a one-dimensional search over $\lambda\in(0,1)$, solving \eqref{eq:final_SDP} at each value.

\section{Details for case studies}\label{app:Details for Case Studies}
In this section, we present the detailed form and parameters of the systems used in Section~\ref{sec:Case Studies}.

\subsection{Unmanned aerial vehicle}
%For system $\Sigma^{(1)}$, we consider the UAV model from \cite[..]{nelson1998flight}
The matrices $A^{(1)}, B^{(1)}$ used in $\Sigma^{(1)}$ are 
obtained by forward-Euler discretization with sampling period 
$\Delta t = 0.02$~s of the linearized longitudinal model 
of~\cite[Chapter 4.4]{nelson1998flight}. We replace the original state variable $w$ (the body-axis vertical velocity) with the angle of attack $a$, defined as the angle between the velocity vector and the forward body axis. Near the nominal airspeed $V_0$, the two are approximately related by $a=w / V_0$. %The $M_{\dot w}$ terms appearing in \cite[eq.~(4.51)]{nelson1998flight} are set to zero, as is standard practice~\cite[Chapter 4.4]{nelson1998flight}. 
Using representative parameters for a small fixed-wing UAV at $V_0 = 15$~m/s, the resulting continuous-time system matrices are:
\begin{equation*}
A_c^{(1)} = 
\begin{bmatrix}
-0.05  & 3.0    & 0    & -9.81 \\
-0.01  & -2.5   & 1    & 0    \\
0.005  & -12.0  & -1.8 & 0    \\
0      & 0      & 1    & 0
\end{bmatrix},
\;
B_c^{(1)} = 
\begin{bmatrix}
0     & 2.0   \\
-0.4  & -0.08 \\
-18.0 & 0.3   \\
0     & 0
\end{bmatrix},
\end{equation*}
and $A^{(1)} = I + \Delta t\,A_c^{(1)}$, $B^{(1)} = \Delta t\,B_c^{(1)}$. 
The output matrix $C^{(1)}$ reflects four onboard sensors 
and is given by:
\begin{equation*}
C^{(1)} = 
\begin{bmatrix}
0 & 1 & 0 & 0 \\
0 & 0 & 1 & 0 \\
1 & 0 & 0 & 0 \\
0 & 0 & 0 & 1
\end{bmatrix}.
\end{equation*}

The matrices $A^{(2)}, B^{(2)}$ used in $\Sigma^{(2)}$ are 
obtained by the same forward-Euler discretization of a 
continuous-time model that augments $\Sigma^{(1)}$ with two additional control surfaces. Each surface follows the second-order spring mass damper dynamics 
of~\cite[Chapter 4.2]{nelson1998flight}:
\begin{equation*}
    I_c\,\dot{\omega}_{\delta_i} = 
    -k_c\,\delta_i - d_c\,\omega_{\delta_i} + K_{a,i}\,a 
    + K_{q,i}\,\omega_\theta + \tau_i,
\end{equation*}
where $I_c = 0.008$ is the surface's moment of inertia, $k_c = 5.0$ the 
spring constant, and $d_c = 0.15$ the damping coefficient. The coupling between the body states and the new surfaces is 
modeled analogously to the elevator coupling 
in~\cite[Chapter 4.4]{nelson1998flight}. A deflection $\delta_i$ 
enters the UAV prototype equations for $a$ and $\omega_\theta$ through 
coefficients $(Z_{\delta_1}, Z_{\delta_2}) = (-0.15, -0.45)$ 
and $(M_{\delta_1}, M_{\delta_2}) = (-10.0, -5.0)$, while the 
body states $a$ and $\omega_\theta$ feed back into the surface 
equations via $(K_{a,1}, K_{a,2}) = (3.0, 2.0)$ and 
$(K_{q,1}, K_{q,2}) = (0.5, 0.8)$. The asymmetric values for 
$i=1,2$ encode the distinct mounting of the two surfaces. At the nominal operating point of $\delta_i = 0$, the body 
dynamics of $\Sigma^{(2)}$ coincide with those of $\Sigma^{(1)}$.
The resulting continuous-time system matrices take the following form:
\begin{equation*}
A_c^{(2)} = 
\begin{bmatrix}
A_c^{(1)} & A_{bs} \\[2pt]
A_{sb}    & A_{ss}
\end{bmatrix},
\quad
B_c^{(2)} = 
\begin{bmatrix}
B_c^{(1)} & 0 \\[2pt]
0         & B_{ss}
\end{bmatrix},
\end{equation*}
with:
\begin{equation*}
A_{bs} = 
\begin{bmatrix}
0     & 0     & 0 & 0 \\
-0.15 & -0.45 & 0 & 0 \\
-10.0 & -5.0  & 0 & 0 \\
0     & 0     & 0 & 0
\end{bmatrix},
\;
A_{ss} = 
\begin{bmatrix}
0    & 0    & 1      & 0 \\
0    & 0    & 0      & 1 \\
-625 & 0    & -18.75 & 0 \\
0    & -625 & 0      & -18.75
\end{bmatrix},
\end{equation*}
\begin{equation*}
A_{sb} = 
\begin{bmatrix}
0 & 0   & 0    & 0 \\
0 & 0   & 0    & 0 \\
0 & 375 & 62.5 & 0 \\
0 & 250 & 100  & 0
\end{bmatrix},
\quad
B_{ss} = 
\begin{bmatrix}
0   & 0   \\
0   & 0   \\
125 & 0   \\
0   & 125
\end{bmatrix},
\end{equation*}
and $A^{(2)} = I + \Delta t\,A_c^{(2)}$, $B^{(2)} = \Delta t\,B_c^{(2)}$, 
and $C^{(2)} = \begin{bmatrix} C^{(1)} & 0_{4 \times 4}\end{bmatrix}$.

\subsection{Hexacopter}
For system $\Sigma^{(1)}$, we use the linearization at hover of the 
Newton--Euler model in~\cite{sabatino2015quadrotor,pounds2010modelling}, 
discretized via forward-Euler with sampling period $\Delta t 
= 0.02$~s. The quadcopter has mass $m^{(1)} = 0.468$, moment of inertia 
$I^{(1)} = 4.9 \cdot 10^{-3}$ about both roll and pitch 
axes, arm length $L^{(1)} = 0.225$, rotational drag 
coefficients $d_p = d_q = 0.012$, and vertical drag 
coefficient $d_z = 0.1$. This yields the continuous-time system matrices:
\begin{equation*}
A_c^{(1)} = 
\begin{bmatrix}
0 & 0 & 1     & 0     & 0 & 0 \\
0 & 0 & 0     & 1     & 0 & 0 \\
0 & 0 & -2.47 & 0     & 0 & 0 \\
0 & 0 & 0     & -2.47 & 0 & 0 \\
0 & 0 & 0     & 0     & 0 & 1 \\
0 & 0 & 0     & 0     & 0 & -0.21
\end{bmatrix},\,
B_c^{(1)} = 
\begin{bmatrix}
0     & 0     & 0      & 0      \\
0     & 0     & 0      & 0      \\
0     & 46.34 & 0      & -46.34 \\
46.34 & 0     & -46.34 & 0      \\
0     & 0     & 0      & 0      \\
2.14  & 2.14  & 2.14   & 2.14
\end{bmatrix},
\end{equation*}
and their discretized counterparts $A^{(1)} = I + \Delta t\,A_c^{(1)}$, $B^{(1)} = \Delta 
t\,B_c^{(1)}$. The output matrix: 
\begin{equation*}
C^{(1)} := 
\begin{bmatrix}
0  & 0 & 1 & 0 & 0 & 0 \\
0  & 0 & 0 & 1 & 0 & 0 \\
-1 & 0 & 0 & 0 & 0 & 0 \\
0  & 1 & 0 & 0 & 0 & 0 \\
0  & 0 & 0 & 0 & 1 & 0
\end{bmatrix}
\end{equation*}
reflects five onboard sensors, where the first two rows measure 
the quadcopter's angular rates $\omega_\phi^{(1)}, 
\omega_\theta^{(1)}$, the next two measure $-\phi^{(1)}$ and 
$\theta^{(1)}$, and the last measures 
the altitude $z^{(1)}$.

For system $\Sigma^{(2)}$, the matrices $A^{(2)}, B^{(2)}$ are 
obtained by the same forward-Euler discretization applied to 
a continuous-time model that follows the Newton--Euler framework 
of~\cite{sabatino2015quadrotor}, adapted to a six-motor 
geometry. The hexacopter has mass $m^{(2)} = 0.88$, moment 
of inertia $I^{(2)} = 0.01$, arm length $L^{(2)} = 0.26$, 
rotational drag coefficients $d_p = d_q = 0.015$, and vertical 
drag coefficient $d_z = 0.12$.
The camera payload tilt angles $\alpha_1, \alpha_2$ obey the second-order 
spring--mass--damper equations 
of~\cite[Chapter 4.2]{nelson1998flight}:
\begin{align*}
    I_g\,\dot{\omega}_{\alpha_1} &= 
    -k_c\,(\alpha_1 + \phi^{(2)}) - d_c\,\omega_{\alpha_1} + \tau_1, \\
    I_g\,\dot{\omega}_{\alpha_2} &= 
    -k_c\,(\alpha_2 + \theta^{(2)}) - d_c\,\omega_{\alpha_2} + \tau_2,
\end{align*}
where $I_g = 5 \cdot 10^{-4}$, $k_c = m_p\,g\,\ell = 1.47 \cdot 
10^{-3}$ with $m_p = 0.03$ and $\ell = 5 \cdot 10^{-3}$, and 
$d_c = 0.04$. The sum $\alpha_i + \phi^{(2)}$ (resp.\ $\alpha_i 
+ \theta^{(2)}$) appears because gravity acts on the absolute 
inclination of the camera. Conversely, the hexacopter receives 
a torque $-\tau_i$ from each camera input by Newton's third law, 
together with a damping reaction $d_c\,\omega_{\alpha_i}$ 
obtained by force balance at the camera attachment. This gives 
the block form:
\begin{equation*}
A_c^{(2)} = 
\begin{bmatrix}
A_{hh} & A_{hc} \\[2pt]
A_{ch} & A_{cc}
\end{bmatrix},
\quad
B_c^{(2)} = 
\begin{bmatrix}
B_{hh} & B_{hc} \\[2pt]
0      & B_{cc}
\end{bmatrix},
\end{equation*}
where the subscript $h$ refers to hexacopter states and $c$ to camera states, with:
\begin{equation*}
A_{hh} = 
\begin{bmatrix}
0 & 0 & 1    & 0    & 0 & 0 \\
0 & 0 & 0    & 1    & 0 & 0 \\
0 & 0 & -1.5 & 0    & 0 & 0 \\
0 & 0 & 0    & -1.5 & 0 & 0 \\
0 & 0 & 0    & 0    & 0 & 1 \\
0 & 0 & 0    & 0    & 0 & -0.14
\end{bmatrix},
A_{cc} = 
\begin{bmatrix}
0     & 0     & 1   & 0   \\
0     & 0     & 0   & 1   \\
-2.94 & 0     & -80 & 0   \\
0     & -2.94 & 0   & -80
\end{bmatrix},
\end{equation*}
\begin{equation*}
A_{hc} = 
\begin{bmatrix}
0 & 0 & 0 & 0 \\
0 & 0 & 0 & 0 \\
0 & 0 & 4 & 0 \\
0 & 0 & 0 & 4 \\
0 & 0 & 0 & 0 \\
0 & 0 & 0 & 0
\end{bmatrix},
\;
A_{ch} = 
\begin{bmatrix}
0     & 0     & 0 & 0 & 0 & 0 \\
0     & 0     & 0 & 0 & 0 & 0 \\
-2.94 & 0     & 0 & 0 & 0 & 0 \\
0     & -2.94 & 0 & 0 & 0 & 0
\end{bmatrix},
\end{equation*}
\begin{equation*}
B_{hh} = 
\begin{bmatrix}
0    & 0     & 0     & 0    & 0      & 0      \\
0    & 0     & 0     & 0    & 0      & 0      \\
0    & 22.52 & 22.52 & 0    & -22.52 & -22.52 \\
26   & 13    & -13   & -26  & -13    & 13     \\
0    & 0     & 0     & 0    & 0      & 0      \\
1.14 & 1.14  & 1.14  & 1.14 & 1.14   & 1.14
\end{bmatrix},
\end{equation*}
\begin{equation*}
B_{hc} = 
\begin{bmatrix}
0    & 0    \\
0    & 0    \\
-100 & 0    \\
0    & -100 \\
0    & 0    \\
0    & 0
\end{bmatrix},
\;
B_{cc} = 
\begin{bmatrix}
0    & 0    \\
0    & 0    \\
2000 & 0    \\
0    & 2000
\end{bmatrix}.
\end{equation*}
The off-diagonal blocks follow directly from the coupled equations, which are obtained by expressing the camera restoring torque in terms of the absolute inclination $\alpha_i + \phi_i^{(2)}$ and by applying Newton’s third law to include the joint damping and input torques in the hexacopter angular dynamics after normalization by the corresponding inertias:
\begin{align*}
\dot{\omega}_{\alpha_1} &= -\tfrac{k_c}{I_g}\alpha_1
- \tfrac{k_c}{I_g}\phi^{(2)} - \tfrac{d_c}{I_g}\omega_{\alpha_1}
+ \tfrac{1}{I_g}\tau_1, \\
\dot{\omega}_{\alpha_2} &= -\tfrac{k_c}{I_g}\alpha_2
- \tfrac{k_c}{I_g}\theta^{(2)} - \tfrac{d_c}{I_g}\omega_{\alpha_2}
+ \tfrac{1}{I_g}\tau_2, \\
\dot{\omega}_{\phi}^{(2)} &= -\tfrac{d_p}{I^{(2)}}\omega_\phi^{(2)}
+ \tfrac{d_c}{I^{(2)}}\omega_{\alpha_1}
- \tfrac{1}{I^{(2)}}\tau_1 + (B_{hh}u)_3, \\
\dot{\omega}_{\theta}^{(2)} &= -\tfrac{d_q}{I^{(2)}}\omega_\theta^{(2)}
+ \tfrac{d_c}{I^{(2)}}\omega_{\alpha_2}
- \tfrac{1}{I^{(2)}}\tau_2 + (B_{hh}u)_4,
\end{align*}
where $(B_{hh}u)_3$ and $(B_{hh}u)_4$ denote the third and fourth components of the vector $B_{hh}u$, corresponding to the $\phi$ and $\theta$ angular dynamics, respectively. The terms $-(k_c/I_g)\phi^{(2)} := -2.94\phi^{(2)}$ and $-(k_c/I_g)\theta^{(2)} := -2.94\theta^{(2)}$ determine the nonzero entries of $A_{ch}$. The terms $(d_c/I^{(2)})\omega_{\alpha_i} := 4\omega_{\alpha_i}$ determine the nonzero entries of $A_{hc}$. Finally, the terms $-(1/I^{(2)})\tau_i := -100\tau_i$ determine the nonzero entries of $B_{hc}$.

\begin{comment}
\section{ACKNOWLEDGMENTS}

\end{comment}

\end{document}